# Exploiting the Hidden Capacity of MMC Through Accurate Quantification of Modulation Indices

Qianhao Sun, *Member, IEEE,* Jingwei Meng, Ruofan Li, Mingchao Xia, *Senior Member, IEEE,* Qifang Chen, *Member, IEEE,* Jiejie Zhou, Meiqi Fan, and Peiqian Guo

***Abstract* — The modular multilevel converter (MMC) has become increasingly important in voltage-source converter-based high-voltage direct current (VSC-HVDC) systems. Direct and indirect modulation are widely used as mainstream modulation techniques in MMCs. However, due to the challenge of quantitatively evaluating the operation of different modulation schemes, the academic and industrial communities still hold differing opinions on their performance. To address this controversy, this paper employs the state-of-the-art computational methods and quantitative metrics to compare the performance among different modulation schemes. The findings indicate that direct modulation offers superior modulation potential for MMCs, highlighting its higher ac voltage output capability and broader linear *PQ* operation region. Conversely, indirect modulation is disadvantaged in linear modulation, which indicates inferior output voltage capability. Furthermore, this paper delves into the conditions whereby direct and indirect modulation techniques become equivalent in steady-state. The study findings suggest that the modulation capability of direct modulation is the same as that of indirect modulation in steady-state when additional controls, including closed-loop capacitor voltage control and circulating current suppression control (CCSC), are simultaneously active. Simulation and experiments verify the correctness and validity.**

## NOMENCLATURE

| | |
|---|---|
| $U_{dcN}$ | Nominal dc voltage. |
| $U_{capN}$ | Rated submodule capacitor voltage. |
| $U_{cap_ref}$ | Preset target value of capacitor dc voltage. |
| $U_{cap}$ | Actual value of capacitor dc voltage. |
| $P_N$ | Rated active power capacity. |
| $Q_{max}$ | Maximum reactive power output. |
| $U_{AVC}$ | Valve-side ac voltage. |
| $U_{ref1}$ | RMS of the fundamental-frequency reference. |
| $U_{ref2}$ | RMS of the second-harmonic reference. |
| $U_{conv1}$ | RMS of required fundamental-frequency ac voltage output. |
| $U_{MS\text{-}ACV}$ | Maximum selectable valve-side ac voltage under premise of whole-operating-range linear modulation. |
| $u^{p,n}_{measured}$ | Instantaneous capacitor voltage from measurement |
| $u^{p,n}_{estimated}$ | Instantaneous capacitor voltage from estimation. |
| $u^{p,n}_{calculated}$ | Instantaneous capacitor voltage based on calculation. |
| $u_{p,n}$ | Arm output voltage |
| $u_{com}$ | Common mode of upper and lower arm voltage |
| $u_{diff}$ | Differential mode of upper and lower arm voltage |
| $\Delta\tilde{u}_{cap}$ | Ripple component of capacitor voltage |
| $\Delta\overline{U}_{cap}$ | DC deviation of capacitor voltage |
| $M_{conv1}$ | Amplitude modulation index of the required ac output voltage. |
| $M_{ref1}$ | Amplitude modulation index of the fundamental-frequency reference. |
| $M_{ref2}$ | Amplitude modulation index of the second-harmonic reference. |
| $\delta_{conv1}$ | Phase of required fundamental-frequency ac voltage output. |
| $\delta_{ref1}$ | Phase of the fundamental-frequency reference. |
| $\delta_{ref2}$ | Phase of the second-harmonic reference. |
| $N$ | Number of submodules in an arm. |
| $C_d$ | Submodule capacitance. |
| $c_1$ | Converter coefficient. |
| $f_{RWF}$ | Reference waveform function (RWF). |
| $F_{peak}$ | Peak value of RWF. |
| $F_{valley}$ | Valley value of RWF. |
| $\Delta F_{margin}$ | Linear modulation margin of RWF. |
| $e_{d_ref}$ | DC reference value generated to regulate capacitor voltage to the target. |
| $e_{cir}$ | Energy balance control item to achieve asymptotic stability |
| $E_{req}$ | Required energy storage requirement. |
| $I_{cir}$ | RMS of the double-frequency circulating current |
| $\theta_{cir}$ | Phase of the second-harmonic circulating current. |
| $k_{cir}$ | Amplitude index of the double-frequency circulating current |
| $h$ | Factor evaluating deviating level of dc component of RWF |
| $I_{ac}$ | RMS of ac current output. |
| $I_{dc}$ | DC link current. |
| $i_{p,n}$ | Arm current. |
| $T_1$ | Fundamental-frequency period. |
| $W_{p,n}$ | Instantaneous arm energy stored |
| $W_{rip}$ | Ripple component of arm energy stored. |
| $W_0$ | DC component of arm energy stored. |
| $X_T$ | Leakage reactance of the transformer |
| $X_{arm}$ | Arm reactance |
| $X_{eq}$ | Equivalent interface reactance |
| $\varphi$ | Power factor angle. |

# I. INTRODUCTION

Modular multilevel converters (MMCs) are widely regarded as a highly promising technology [1-2], given their numerous advantages, including the ease of implementing a high number of levels, excellent harmonic performance, and low losses [3-5]. In particular, this technology has emerged as a competitive solution for voltage-source converter-based high-voltage direct current (VSC-HVDC) transmission systems [6-8].

Similar to existing VSCs, the performance of an MMC is closely linked to the modulation technique utilized [9].

On the one hand, as a replacement for the conventional two-level VSCs in the high-voltage domain, the MMC can employs direct modulation, a technique inherited from its predecessor. However, the physical separation of the six arms in MMCs causes significant power fluctuations in each arm, leading to stored energy variations and subsequent ripples on submodule (SM) capacitor voltages [10]. Consequently, the primary limitation of direct modulation technique is the existence of circulating currents [11-12], which increase converter power losses and device ratings [13-15].

On the other hand, in contrast to direct modulation, indirect modulation that produces precisely the required voltage has been studied, particularly in Europe. A closed-loop indirect modulation has been proposed in [16], requiring the integration of capacitor voltage control and an additional energy balance controller among bridge arms to ensure stability [17]. To reduce the effects of measurement, an open-loop indirect modulation proposed in [18]-[19] uses estimated capacitor voltages to take place of measured capacitor voltages. In this methodology, capacitor voltages are derived through the resolution of equations that model the converter's underlying dynamics. This approach offers several advantages, such as the elimination of voltage sensors, the absence of communication requirements, and the guarantee of global asymptotic stability. Nevertheless, a key drawback lies in the necessity for accurately extracting the genuine parameters that accurately represent the system dynamics, as the precision of these estimates directly impacts the accuracy of the capacitor voltage calculation.

According to the abovementioned analyses, the qualitative advantages and disadvantages of direct and indirect modulation in MMCs are clear now based on the contributions of previous works. However, to the best of the authors' knowledge, there has been no quantitative comparison reports or results for direct and indirect modulation thus far. The main reason for this is that the precise quantitative evaluation of modulation performance for direct modulation has consistently posed a significant challenge for researchers due to the deviation between the actual voltage and the reference voltage [20]. Unlike indirect modulation, which provides precise voltage outputs, the evaluation of modulation capability for direct modulation is pretty complex. Given this research gap, a quantitative comparison of direct and indirect modulation in MMCs is crucial and necessary for the in-depth optimization of MMC design and operation. In particular, some qualitative assertions about modulation performance have not received sufficient support in terms of quantitative calculations. For example, based on qualitative analyses, reference [21] has concluded that "indirect modulation, based on the instantaneous voltage of arm capacitors, can utilize capacitor voltage ripple to extend output voltage capability." However, the research findings of this paper, based on a quantitative comparison, do not align with the viewpoint. Moreover, as far as the authors are aware, despite the second-harmonic circulating current and dc capacitor voltage deviation being the primary distinctions between direct and indirect modulation, a lack of quantitative analysis has prevented several key issues from being investigated thoroughly until now:

1) There is a lack of quantitative evaluation and comparison among different modulation techniques. Consequently, the linear modulation capability of different modulation techniques has not been clearly compared and discussed quantitatively in the existing public literature.

2) The effects of additional controls such as CCSC and capacitor voltage control on linear modulation have not been systematically studied. It is well-known that these controls have negative effects on linear modulation, but quantitative results are not available.

3) Different modulations have different characteristics, and these differences will impact the optimization design of MMCs, such as the rated valve-side ac voltage and the SM capacitance used. However, to the best of the authors' knowledge, related studies have not yet been reported.

4) Given the differences between direct and indirect modulation, it remains unclear whether one can be converted into the other. If such a conversion is possible, the conditions for it need to be determined.

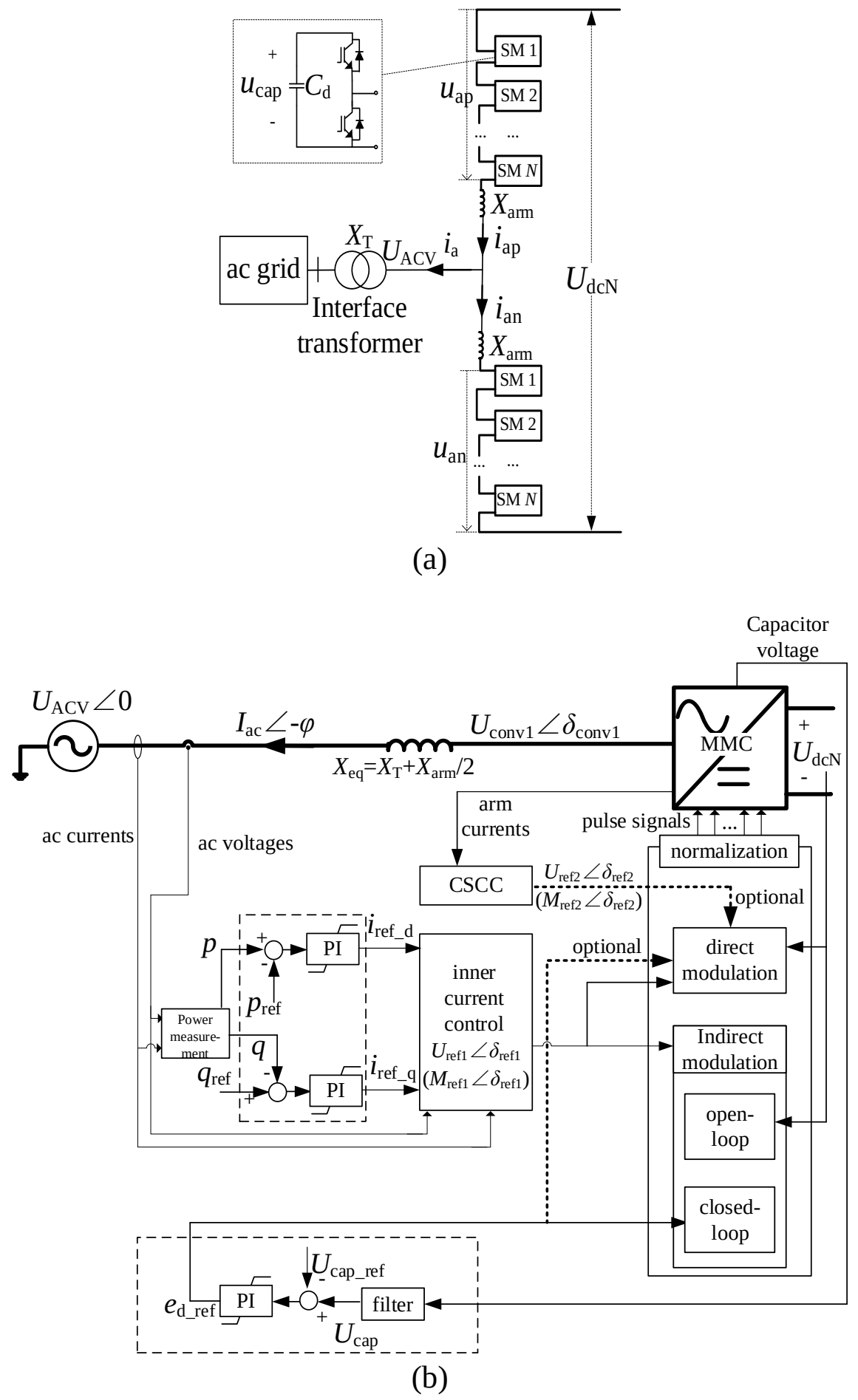

Fig. 1. Diagram of MMC connected to an ac grid: (a) single-line diagram of an MMC (b) equivalent circuit and controller diagram

To further understand the modulation of MMCs, the detailed quantitative comparison of different modulations is made in this paper, and several key conclusions are first presented.

The first contribution of this paper is the development of a quantitative comparison framework for modulation techniques of MMCs, which enables the quantitative comparison of different modulation schemes. This work is the result of incorporating and expanding upon several recent developments in the theory of direct modulation [20], [22]-[23]. In [20] and [22], the precise calculation method of the reference wave function (RWF) is proposed for direct modulation, which is used as a quantitative comparison tool for reflecting the modulation performance in this paper. Meanwhile, in [22], quantifiable metric called linear modulation range is introduced, which is used in this paper to quantitatively assess the modulation capability and the effect of design. At last, [23] enhances the theory of capacitor voltage regulation in direct modulation and proposed an improved direct modulation based on capacitor voltage closed-loop control and circulating current suppression control (CCSC). (To distinguish the work of [23] from the conventional direct modulation, this paper refers to it as improved direct modulation). According to this paper, the improved direct modulation will provide the steady-state equivalent conditions between direct and indirect modulation. Other contributions of this paper are concluded as:

1) The detailed comparisons of RWF among direct, indirect and improved direct modulation under various typical operating conditions of MMCs, which is based on a precise quantitative calculation analysis, are provided in this paper. Consequently, the fundamental differences of direct and indirect modulation can be comprehensively explained and examined. Meanwhile, the operation consistency of RWF between indirect and improved direct modulation is also examined by these quantitative comparisons.

2) The comprehensive effects of different modulation techniques on linear modulation margin, capacitor voltage waveform and linear *PQ* region in which linear modulation is always satisfied, are quantitatively compared in detail. In particular, a precise calculation method for determining the linear *PQ* operating region indicating modulation capability is proposed to facilitate this comparison.

3) The equivalence not only in RWF but also in linear modulation margin, capacitor voltage and linear *PQ* region between indirect modulation and improved direct modulation has been demonstrated through detailed comparisons. Consequently, the conditions under which direct and indirect modulation are equivalent in steady-state have been identified.

4) The effects of different modulation on parameter design, including the rated valve-side ac voltage and energy storage (i.e., capacitance used), are considered for a given *PQ* operating range. In general, the valve-side ac voltage should be as high as possible to decrease output current and, in turn, reduce capacitor voltage ripples and the capacitance used with the same peak capacitor voltage. However, a higher valve-side voltage also increases the likelihood of overmodulation. Therefore, the evaluation of different modulation techniques on parameter design should balance these competing factors.

The remainder of this paper is organized as follows: Section II briefly reviews the principles of different modulation schemes. Section III proposes a framework for the quantitative comparison of different modulation techniques. Section IV presents a precise quantitative calculation for RWF using different modulation schemes, and a detailed comparison of RWF, which directly reflects the similarities and differences of various modulation techniques, is also made. In Section V, detailed comparisons of key metrics and design parameters are made in this section to clearly distinguish the pros and cons of each. Section VI verifies the analyses through simulation and experiment. Meanwhile, the evaluation and comparison of voltage THD and losses are conducted. Section VII discusses the conclusions.

## II. Principles of Modulation Techniques in MMCs

### *A. Modulation mechanism of MMC*

Fig. 1(a) shows the topology of an MMC. The dc side of the MMC is connected to the dc line, and the nominal dc voltage is denoted as $U_{\mathrm{dcN}}$. Each arm contains $N$ series-connected SMs, and the rated capacitor voltage is expressed as

$$U_{\mathrm{capN}} = \frac{U_{\mathrm{dcN}}}{N} \tag{1}$$

The ac side of the MMC is connected to an ac power grid via an interface transformer. The rated valve-side ac voltage of the interface transformer is denoted using $U_{\mathrm{ACV}}$.

Fig.1 (b) shows the equivalent circuit and the modulation mechanism of MMC. The ac side is equivalent to be connected to an ac voltage source of $U_{\mathrm{ACV}}$ via an equivalent reactance $X_{\mathrm{eq}}$, which is determined by the leakage reactance of the transformer $X_{\mathrm{T}}$ and the arm reactance $X_{\mathrm{arm}}$ of MMC.

For indirect and direct modulation, the ac-side controller is the same. The outer loop in the ac side ensures active (or dc link voltage) and reactive power meet the target values, which generates fundamental-frequency reference voltage $U_{\mathrm{ref1}}\angle\delta_{\mathrm{ref1}}$.

However, the situation is distinct for the dc-side controller. For example, in closed-loop indirect modulation, a closed-loop control is employed to regulate the capacitor voltage strictly, and a dc reference value denoted as $e_{\mathrm{d_ref}}$ is generated to ensure that the capacitor voltage $U_{\mathrm{cap}}$ follows the target value $U_{\mathrm{cap_ref}}$ precisely. Conversely, for direct modulation and open-loop indirect modulation, the closed-loop capacitor voltage control is usually deemed as unnecessary, although in some special applications, such as the improved direct modulation in [23], capacitor voltage control and CCSC are still required.

Although the implement of controller differs, the following normalization is the key for various modulation.

### *B. Normalization of different modulation techniques in MMCs*

To express the outcome of normalization, RWF is used in this paper. The time-domain expression of RWF is denoted as $f_{\mathrm{RWF}}$. Normalization is essentially a division operation, where the numerator comprises the summation of the ac and dc voltage references, while the denominator is the summation of the capacitor voltages. The normalization differences for various modulation schemes directly reflect in the differences of the definition expressions. To illustrate these differences, the definitions of $f_{\mathrm{RWF}}$ for different modulation techniques are presented as follows. For simplifying the analyses, this paper only considers the case of phase A, as the analyses for phases B and C are similar.

*B1. RWF definition expression of indirect modulation*

*1) Closed-loop indirect modulation*

In this modulation, the upper and lower arm RWFs are controlled by the two complementary sinusoidal reference waveforms, as given by

$$f_{\mathrm{RWF},\,p} = \frac{\frac{e_{\mathrm{d_ref}}}{2} - \sqrt{2}U_{\mathrm{ref1}}\sin(\omega t + \delta_{\mathrm{ref1}}) + e_{\mathrm{cir}}}{Nu^{p}_{\mathrm{measured}}} \tag{2a}$$

$$f_{\mathrm{RWF},\,n} = \frac{\frac{e_{\mathrm{d_ref}}}{2} + \sqrt{2}U_{\mathrm{ref1}}\sin(\omega t + \delta_{\mathrm{ref1}}) + e_{\mathrm{cir}}}{Nu^{n}_{\mathrm{measured}}} \tag{2b}$$

where $u^{p,\,n}_{\mathrm{measured}}$ stand for instantaneous capacitor voltage of upper and lower arm, respectively from measurement. $e_{\mathrm{cir}}$ is the energy balance control item to achieve asymptotic stability, and the value of $e_{\mathrm{cir}}$ approaches zero during the steady-state operation.

*2) Open-loop indirect modulation*

In this modulation, the upper and lower arm RWFs are:

$$f_{\mathrm{RWF},\,p} = \frac{\frac{U_{\mathrm{dcN}}}{2} - \sqrt{2}U_{\mathrm{ref1}}\sin(\omega t + \delta_{\mathrm{ref1}})}{Nu^{p}_{\mathrm{estimated}}} \tag{3a}$$

$$f_{\mathrm{RWF},\,n} = \frac{\frac{U_{\mathrm{dcN}}}{2} + \sqrt{2}U_{\mathrm{ref1}}\sin(\omega t + \delta_{\mathrm{ref1}})}{Nu^{n}_{\mathrm{estimated}}} \tag{3b}$$

In the case of open-loop modulation, the instantaneous values $u^{p,\,n}_{\mathrm{estimated}}$ are obtained via estimation.

*B2. RWF definition expression of direct modulation*

In this modulation, RWFs of upper and lower arms are also controlled by the two complementary sinusoidal reference waveforms, as given by

$$f_{\mathrm{RWF},\,p} = \frac{\frac{U_{\mathrm{dcN}}}{2} - \sqrt{2}U_{\mathrm{ref1}}\sin(\omega t + \delta_{\mathrm{ref1}})}{NU_{\mathrm{cap_ref}}} \tag{4a}$$

$$f_{\mathrm{RWF},\,n} = \frac{\frac{U_{\mathrm{dcN}}}{2} + \sqrt{2}U_{\mathrm{ref1}}\sin(\omega t + \delta_{\mathrm{ref1}})}{NU_{\mathrm{cap_ref}}} \tag{4b}$$

The numerator of the RWF remains the same as that of open-loop indirect modulation. However, for the denominator, the summation of instantaneous capacitor voltages is substituted with the summation of all the capacitor voltage references, $NU_{\mathrm{cap_ref}}$. On the one hand, this substitution enables direct modulation to eliminate the need for communication related to transmitting measured voltage values and energy balance control, when compared to the closed-loop indirect modulation. And it also avoids the complicated estimation of capacitor voltage required by open-loop indirect modulation. On the other hand, the use of $NU_{\mathrm{cap_ref}}$ introduces the well-known harmonic circulating current, which increases the losses of the converter. Meanwhile, the absence of capacitor voltage control results in the deviation of the dc value of the capacitor voltage from the reference value $U_{\mathrm{cap_ref}}$ [24]. Therefore, to address the above two issues, improved direct modulation is introduced in [23]. This modulation technique requires the use of CCSC and closed-loop capacitor voltage control, while still using $NU_{\mathrm{cap_ref}}$ as the denominator during the normalization.

Then, for the normalization of improved direct modulation, $U_{\mathrm{dcN}}$ in (4a-4b) should be substituted with the $e_{\mathrm{d_ref}}$ for capacitor voltage control and the injected second-harmonic reference voltage $U_{\mathrm{ref2}}\angle\delta_{\mathrm{ref2}}$ for CCSC also should be taken into account. By doing so, the RWF definition expression based on improved direct modulation can be derived as:

$$f_{\mathrm{RWF},\,p} = \frac{\frac{e_{\mathrm{d_ref}}}{2} - \sqrt{2}U_{\mathrm{ref1}}\sin(\omega t + \delta_{\mathrm{ref1}}) + \sqrt{2}U_{\mathrm{ref2}}\sin(2\omega t + \delta_{\mathrm{ref2}})}{NU_{\mathrm{cap_ref}}} \tag{5a}$$

$$f_{\mathrm{RWF},\,n} = \frac{\frac{e_{\mathrm{d_ref}}}{2} + \sqrt{2}U_{\mathrm{ref1}}\sin(\omega t + \delta_{\mathrm{ref1}}) + \sqrt{2}U_{\mathrm{ref2}}\sin(2\omega t + \delta_{\mathrm{ref2}})}{NU_{\mathrm{cap_ref}}} \tag{5b}$$

Based on (2a) - (5b), it can be observed that the definition expression of RWF is different for different modulation schemes. However, the differences in the definition expressions of RWF in (2a) - (5b) are only superficial and qualitative, and a quantitative comparison of them requires further calculations.

## III. Establishment of Quantitative Comparison Framework for Different Modulation Schemes

A theoretical framework for how to achieve quantitative comparison of different modulation techniques in MMCs is established in this section, and several important quantifiable metrics and parameters design in the proposed framework are also explained.

As shown in Fig.2, the precise expression of $f_{\mathrm{RWF}}$ is used as the analytical foundation. Meanwhile, the detailed quantitative evaluation metrics of RWF for different modulation schemes are also concluded in Fig.2. Then comprehensive effects of modulation are compared through quantifiable metrics to reveal different capabilities and performances of different modulation. In the proposed framework, the linear modulation margin $\Delta F_{\mathrm{margin}}$ and capacitor voltages are chosen for comparison at a specific point, while a linear *PQ* operating region is utilized for comparison based on given design parameters. Furthermore, considering a required *PQ* operating range and peak capacitor voltage limit, the impact of various modulation techniques on parameter design is examined, including the maximum selectable valve-side ac voltage $U_{\mathrm{MS\text{-}ACV}}$ (MS-ACV) and the required energy storage $E_{\mathrm{req}}$.

*A. Linear modulation margin*

The linear modulation margin, $\Delta F_{\mathrm{margin}}$, is used to quantitatively evaluate the margin to the boundary of linear modulation constraints for any operating point. Linear modulation for a certain operating point is satisfied if $\Delta F_{\mathrm{margin}}$ is greater than zero. The lower the value of $\Delta F_{\mathrm{margin}}$, the closer the MMC to the boundary condition of overmodulation. The specific calculation method for $\Delta F_{\mathrm{margin}}$ is as follows.

For a HB-MMC, the linear modulation constraint is:

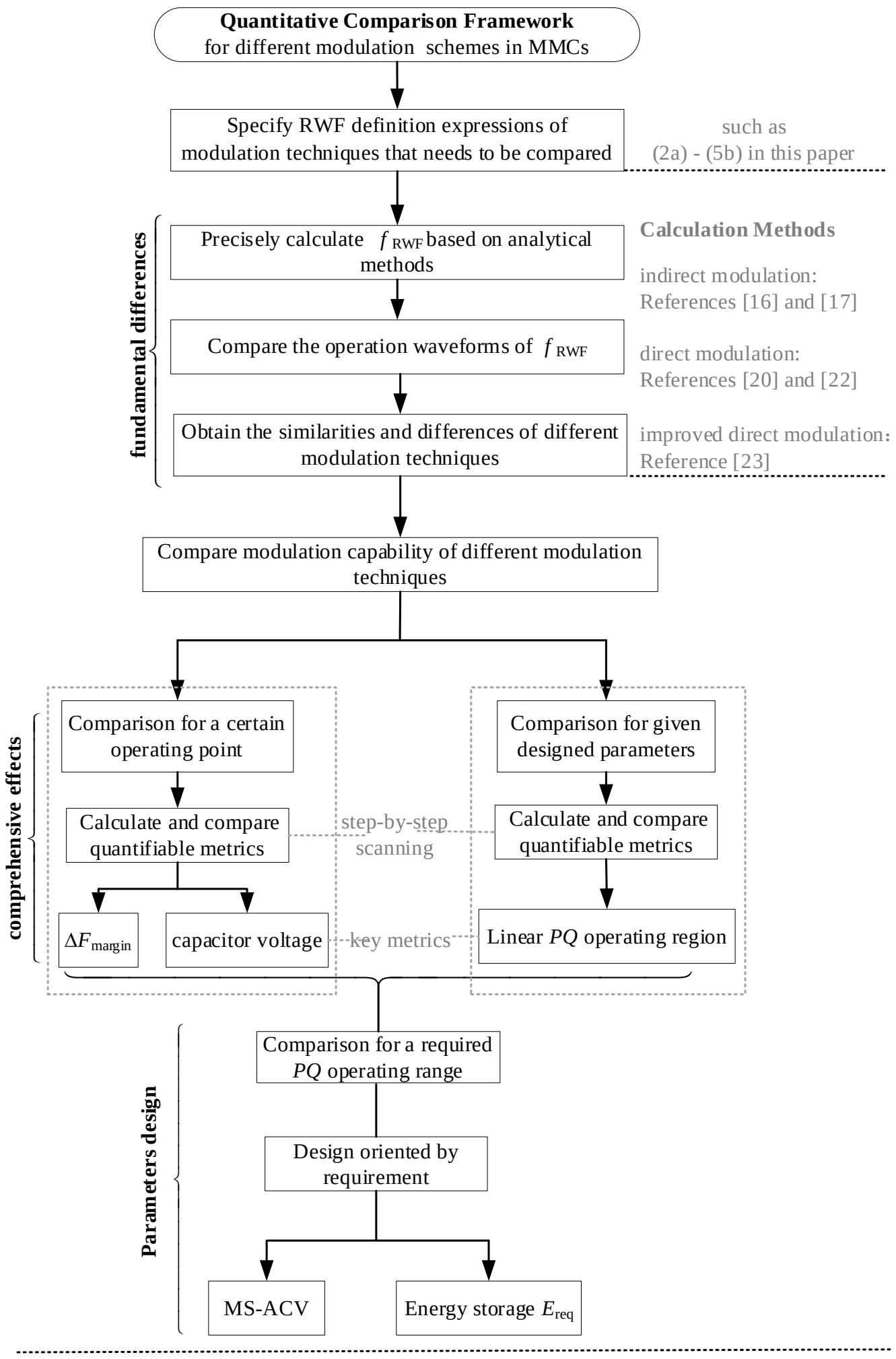

Fig. 2. The quantitative comparison framework for different modulation techniques in MMCs

$$0 \le f_{RWF}(t) \le 1 \tag{6}$$

Assumed the peak and valley values of RWF in a period are denoted as $F_{peak}$ and $F_{valley}$, respectively, and they can be expressed as follows:

$$\begin{cases} F_{peak} = \max\left(f_{RWF}(t)\right) \\ F_{valley} = \min\left(f_{RWF}(t)\right) \end{cases} \quad 0 \le t \le T_1 \tag{7}$$

where $T_1$ is the fundamental-frequency period.

Based on (7), (6) can be transformed into:

$$F_{peak} \le 1 \text{ and } F_{valley} \ge 0 \tag{8}$$

In equation (8), the upper and lower boundary constraints both should be satisfied. Therefore, $\Delta F_{margin}$ is defined as the minimum of the upper and lower margins:

$$\Delta F_{margin} = \min\left[\left(F_{valley} - 0\right),\ \left(1.0 - F_{peak}\right)\right] \tag{9}$$

So, if $\Delta F_{margin} > 0$, the linear modulation is satisfied.

### *B. Capacitor voltages*

Capacitor voltages not only affect output voltage of an arm, but also determine the energy storage requirement. Meanwhile, ripple voltages bring additional voltage stress on devices in SMs. Thus, SM capacitance should be sufficiently large to limit the capacitor voltage peak within an acceptable level, thereby causing the capacitor to take an extremely large proportion in the cost and volume of an SM [25]. So, it is essential to investigate the capacitor voltages under different modulation techniques.

### *C. Linear PQ operating region*

The linear *PQ* operating region for given parameters is an important metric that reflects actual active and reactive power output capabilities of the converter. The precise quantitative calculation of linear *PQ* operating region is complicated, particularly for direct and improved direct modulation due to the excessively intricate analytical process. To address this issue, a linear *PQ* region calculation flowchart for different modulation is also proposed and used in this paper.

### *D. Maximum selectable valve-side ac voltage (MS-ACV)*

The design of the converter necessitates that all operating points within the entire required operating range adhere to the linear modulation constraint, which is defined as whole operating range linear modulation (WOR-LM) in [22]. Reference [22] demonstrates that the selection of the per-unit valve-side voltage of the interface transformer $U^*_{ACV}$ (normalized by $U_{dcN}/2$) is critical in terms of satisfying WOR-LM. A higher valve-side voltage is preferable for minimizing arm currents, losses, and SM capacitance when the grid-side voltage is constant and limited by external conditions. Thus, the rated valve-side $U^*_{ACV}$ should be selected as high as possible for a given rated dc-side voltage and given *PQ* operating range, provided that WOR-LM is satisfied. Then, to quantitatively evaluate the limitation of $U^*_{ACV}$ for different modulation techniques, the maximum selectable valve-side ac voltage (MS-ACV), denoted as $U^*_{MS\text{-}ACV}$, is defined as the maximum per-unit valve-side voltage that can be utilized while still satisfying WOR-LM, and the detailed calculation method can be found in [22]. To evaluate the output capability of ac voltage, the effects of different modulation schemes on MS-ACV should also be investigated.

### *E. Energy storage*

Given that different modulation techniques lead to different MS-ACV and capacitor voltage waveforms, it is crucial to estimate the energy storage (i.e., capacitance used) with the same peak capacitor voltage limit in detail, while accounting for the safe margin left for the devices

Based on the analysis framework of Fig.2, the comparison and conclusions for different modulation schemes are as follows.

## IV. QUANTITATIVE ANALYSES AND COMPARISONS OF RWF

### *A. Analytical calculations of RWF*

### *A1. Indirect modulation*

In steady-state operation, closed-loop and open-loop indirect modulation schemes exhibit similar behaviors, so the dc reference $e_{d_ref}$ approximates to $U_{dcN}$, and the fundamental-frequency reference, $U_{ref1} \angle \delta_{ref1}$, approaches the actual ac output voltage, $U_{conv1} \angle \delta_{conv1}$. Then, both modulation techniques can be expressed in steady-state as follows:

$$f_{\mathrm{RWF},\,p}=\frac{\frac{U_{\mathrm{dcN}}}{2}-\sqrt{2}U_{\mathrm{conv1}}\sin(\omega t+\delta_{\mathrm{conv1}})}{Nu_{\mathrm{calculated}}^{p}} \tag{10a}$$

$$f_{\mathrm{RWF},\,n}=\frac{\frac{U_{\mathrm{dcN}}}{2}+\sqrt{2}U_{\mathrm{conv1}}\sin(\omega t+\delta_{\mathrm{conv1}})}{Nu_{\mathrm{calculated}}^{n}} \tag{10b}$$

To calculate the RWF in indirect modulation, the actual ac output voltage $U_{\mathrm{conv1}}\angle\delta_{\mathrm{conv1}}$ and the calculated capacitor voltage $u_{\mathrm{calculated}}^{p,\,n}$ must be calculated in advance. Using the calculation method illustrated in [16-17], $U_{\mathrm{conv1}}\angle\delta_{\mathrm{conv1}}$ and capacitor voltage $u_{\mathrm{calculated}}^{p,\,n}$ can be precisely calculated as follows.

*1) Actual ac output voltage*

When MMC generates a fundamental-frequency ac current $I_{\mathrm{ac}}\angle-\varphi$, the corresponding fundamental-frequency ac voltage output $U_{\mathrm{conv1}}$ and the phase angle $\delta_{\mathrm{conv1}}$ are:

$$U_{\mathrm{conv1}}=\sqrt{\left(U_{\mathrm{ACV}}+X_{\mathrm{eq}}I_{\mathrm{ac}}\sin\varphi\right)^{2}+\left(X_{\mathrm{eq}}I_{\mathrm{ac}}\cos\varphi\right)^{2}} \tag{11}$$

$$\delta_{\mathrm{conv1}}=\arctan\frac{X_{\mathrm{eq}}I_{\mathrm{ac}}\cos\varphi}{U_{\mathrm{ACV}}+X_{\mathrm{eq}}I_{\mathrm{ac}}\sin\varphi} \tag{12}$$

where $U_{\mathrm{ACV}}$ denotes the valve-side voltage of the MMC; $X_{\mathrm{eq}}$ denotes the equivalent reactance.

Assumed a specific dc voltage $U_{\mathrm{dcN}}$ is given for the MMC, $U_{\mathrm{conv1}}$ is generally represented by the amplitude modulation index. The index is defined as the ratio of the amplitude of the line-to-ground ac voltage to half of the dc voltage:

$$\begin{aligned}M_{\mathrm{conv1}}&=\frac{\sqrt{2}U_{\mathrm{conv1}}}{U_{\mathrm{dcN}}/2}\\&=U_{\mathrm{ACV}}^{*}\sqrt{\left(1+X_{\mathrm{eq}}^{*}I_{\mathrm{ac}}^{*}\sin\varphi\right)^{2}+\left(X_{\mathrm{eq}}^{*}I_{\mathrm{ac}}^{*}\cos\varphi\right)^{2}}\end{aligned} \tag{13}$$

where $U^{*}_{\mathrm{ACV}}$ is the per-unit valve-side voltage normalized by $U_{\mathrm{dcN}}/2$, $I^{*}_{\mathrm{ac}}$ is the per-unit ac current output, and $X^{*}_{\mathrm{eq}}$ is the per-unit equivalent reactance.

*2) Capacitor voltage*

The capacitor voltages of the upper and lower arms in MMCs can be calculated as follows:

$$u_{\mathrm{calculated}}^{p}=\sqrt{\frac{2W_{p}}{NC_{\mathrm{d}}}} \tag{14a}$$

$$u_{\mathrm{calculated}}^{n}=\sqrt{\frac{2W_{n}}{NC_{\mathrm{d}}}} \tag{14b}$$

where $W_{p,\,n}$ is the energy stored in the capacitors of the upper and lower arms; and $C_{\mathrm{d}}$ is the SM capacitance.

Meanwhile, since the energy storage comprises both dc and ripple components, the following equations can be established:

$$W_{p}=W_{0}+W_{\mathrm{rip},\,p}=\frac{1}{2}C_{\mathrm{d}}U_{\mathrm{cap_ref}}^{2}N+\int u_{p}\times i_{p}dt \tag{15a}$$

$$W_{n}=W_{0}+W_{\mathrm{rip},\,n}=\frac{1}{2}C_{\mathrm{d}}U_{\mathrm{cap_ref}}^{2}N+\int u_{n}\times i_{n}dt \tag{15b}$$

where $u_{p}$ ($u_{n}$) and $i_{p}$ ($i_{n}$) denote the output voltage and arm current of the upper (lower) arm, respectively. The ripple energy $W_{\mathrm{rip}}$ can be obtained by integrating the arm power as:

$$\begin{aligned}W_{\mathrm{rip},\,p}&=\int u_{p}\times i_{p}dt\\&=\int\left[\frac{U_{\mathrm{dcN}}}{2}-\sqrt{2}U_{\mathrm{conv1}}\sin(\omega t+\delta_{\mathrm{conv1}})\right]\left[\frac{\sqrt{2}}{2}I_{\mathrm{ac}}\sin\left(\omega t-\varphi\right)+\frac{I_{\mathrm{dc}}}{3}\right]dt\end{aligned} \tag{16a}$$

$$\begin{aligned}W_{\mathrm{rip},\,n}&=\int u_{n}\times i_{n}dt\\&=\int\left[\frac{U_{\mathrm{dcN}}}{2}+\sqrt{2}U_{\mathrm{conv1}}\sin(\omega t+\delta_{\mathrm{conv1}})\right]\left[-\frac{\sqrt{2}}{2}I_{\mathrm{ac}}\sin\left(\omega t-\varphi\right)+\frac{I_{\mathrm{dc}}}{3}\right]dt\end{aligned} \tag{16b}$$

where $I_{\mathrm{dc}}$ is the dc current.

Based on (10a) - (16b), the precise quantitative expression of RWF for indirect modulation can be calculated.

*A2. Direct modulation*

Assuming $U_{\mathrm{cap_ref}}$ is set to $U_{\mathrm{capN}}$, which is commonly adopted in studies and applications, (4a-4b) can be rewritten as:

$$f_{\mathrm{RWF},\,p}=\frac{1}{2}-\frac{M_{\mathrm{ref1}}}{2}\sin(\omega t+\delta_{\mathrm{ref1}}) \tag{17a}$$

$$f_{\mathrm{RWF},\,n}=\frac{1}{2}+\frac{M_{\mathrm{ref1}}}{2}\sin(\omega t+\delta_{\mathrm{ref1}}) \tag{17b}$$

where $M_{\mathrm{ref1}}$ represents the magnitude ratio of the fundamental-frequency reference wave and is expressed as:

$$M_{\mathrm{ref1}}=\frac{\sqrt{2}U_{\mathrm{ref1}}}{U_{\mathrm{dcN}}/2} \tag{18}$$

To precisely calculate the values of $M_{\mathrm{ref1}}$ and $\angle\delta_{\mathrm{ref1}}$ for direct modulation, references [20] and [22] mentioned in the proposed comparison framework point that the capacitor voltage ripple effect should be considered. Then, the detailed calculation is categorized into common and differential modes.

*1) Common mode of upper and lower arms*

The common mode of the upper and lower arms consists of dc and double-frequency component:

$$\begin{aligned}u_{\mathrm{com}}&=f_{\mathrm{RWF},\,p}\times N\times u_{\mathrm{cap},\,p}+f_{\mathrm{RWF},\,n}\times N\times u_{\mathrm{cap},\,n}\\&=N\left(U_{\mathrm{capN}}+\Delta\overline{U}_{\mathrm{cap}}+\Delta\tilde{u}_{\mathrm{cap(2)}}\right)\\&\quad-NM_{\mathrm{ref1}}\sin\left(\omega t+\delta_{\mathrm{ref1}}\right)\left(\Delta\tilde{u}_{\mathrm{cap(1)}}+\Delta\tilde{u}_{\mathrm{cap(3)}}\right)\end{aligned} \tag{19}$$

where $\Delta\tilde{u}_{\mathrm{cap(1)}}$, $\Delta\tilde{u}_{\mathrm{cap(2)}}$, and $\Delta\tilde{u}_{\mathrm{cap(3)}}$ denote the fundamental-frequency, second-harmonic, and third-harmonic components of capacitor voltage, respectively, they are illustrated in appendix A; $\Delta\overline{U}_{\mathrm{cap}}$ represents the dc deviation of capacitor voltage from the reference, which has been reported in [24].

Since the dc component of $u_{\mathrm{com}}$ is equivalent to $U_{\mathrm{dcN}}$, the following equation can be established based on (19):

$$\begin{aligned}\Delta\overline{U}_{\mathrm{cap}}^{*}=&-4c_{1}M_{\mathrm{ref1}}I_{\mathrm{ac}}^{*}\sin\left(\varphi+\delta_{\mathrm{ref1}}\right)\\&-4c_{1}M_{\mathrm{ref1}}^{2}k_{\mathrm{cir}}I_{\mathrm{ac}}^{*}\cos\left(\theta_{\mathrm{cir}}-2\delta_{\mathrm{ref1}}\right)\end{aligned} \tag{20}$$

where $\Delta\overline{U}_{\mathrm{cap}}^{*}$ represents the per-unit value of $\Delta\overline{U}_{\mathrm{cap}}$ normalized by $U_{\mathrm{capN}}$, and $c_{1}$ is the coefficient, which is presented in Appendix A. The amplitude index of the double-frequency circulating current $k_{\mathrm{cir}}$ is defined by

$$k_{\text{cir}} = \frac{I_{\text{cir}}}{I_{\text{ac}}} \tag{21}$$

where $I_{\text{cir}}$ and $\theta_{\text{cir}}$ denote the RMS value and phase of the second-harmonic circulating current, respectively. The detailed expressions of $I_{\text{cir}}$ and $\theta_{\text{cir}}$ are provided in Appendix A.

Since $k_{\text{cir}}$ is negligible, $\Delta\overline{U}^*_{\text{cap}}$ is primarily determined by the first term on the right-hand side of (20). When the MMC only transmits active power, $\Delta\overline{U}^*_{\text{cap}}$ approximates to zero. When the MMC outputs reactive power (i.e., $0<\varphi<\pi$), $\Delta\overline{U}^*_{\text{cap}}$ is negative, indicating that the actual capacitor voltage is lower than the reference. Conversely, when the MMC receives reactive power (i.e., $-\pi<\varphi<0$), $\Delta\overline{U}^*_{\text{cap}}$ is positive, indicating that the actual capacitor voltage is higher than the reference.

*2) Differential mode of upper and lower arms*

The differential mode of the upper and lower arms contains ac fundamental-frequency component and can be obtained as:

$$\begin{aligned} u_{\text{diff}} &= \frac{1}{2}\left( f_{\text{RWF, n}} \times N \times u_{\text{cap, n}} - f_{\text{RWF, p}} \times N \times u_{\text{cap, p}} \right) \\ &= \frac{N}{2}\left(-\Delta\tilde{u}_{\text{cap(1)}} - \Delta\tilde{u}_{\text{cap(3)}}\right) \\ &\quad + \frac{NM_{\text{ref1}}}{2}\sin\left(\omega t + \delta_{\text{ref1}}\right)\left(U_{\text{capN}} + \Delta\overline{U}_{\text{cap}} + \Delta\tilde{u}_{\text{cap(2)}}\right) \end{aligned} \tag{22}$$

Since the fundamental-frequency component in per-unit is always equal to actual ac required output voltage $M_{\text{conv1}}\angle\delta_{\text{conv1}}$, the following equation can be derived:

$$\begin{aligned} M_{\text{conv1}}\sin(\omega t + \delta_{\text{conv1}}) &\equiv M_{\text{ref1}}\sin\left(\omega t + \cos\delta_{\text{ref1}}\right) \\ &+ \underbrace{c_1 I^*_{\text{ac}}(8 - 3M^2_{\text{ref1}})\sin(\varphi + \delta_{\text{ref1}})\sin(\omega t + \delta_{\text{ref1}})}_{\text{direct deviation}} \\ &+ \underbrace{c_1 I^*_{\text{ac}}(8 - 3M^2_{\text{ref1}})\cos(\varphi + \delta_{\text{ref1}})\cos(\omega t + \delta_{\text{ref1}})}_{\text{quadrature deviation}} \\ &\underbrace{+12c_1 M_{\text{ref1}} k_{\text{cir}} I^*_{\text{ac}}\sin\left(\omega t + \theta_{\text{cir}} - \delta_{\text{ref1}}\right) - 4c_1 M^3_{\text{ref1}} k_{\text{cir}} I^*_{\text{ac}}\cos(2\delta_{\text{ref1}} - \theta_{\text{cir}})\sin\left(\omega t + \delta_{\text{ref1}}\right)}_{\text{related with second-harmonic circulating current}} \end{aligned} \tag{23}$$

Considering that $k_{\text{cir}}$ is relatively minor, the items that contains $k_{\text{cir}}$ can be disregarded. In terms of reactive power transmission, $M_{\text{conv1}}\angle\delta_{\text{conv1}}$ primarily depends on $M_{\text{ref1}}\angle\delta_{\text{ref1}}$ and direct deviation. When MMC outputs reactive power, $M_{\text{ref1}}$ is lower than $M_{\text{conv1}}$ , whereas when it receives reactive power, $M_{\text{ref1}}$ is higher than $M_{\text{conv1}}$. In terms of active power transmission, $M_{\text{conv1}}\angle\delta_{\text{conv1}}$ mainly relies on $M_{\text{ref1}}\angle\delta_{\text{ref1}}$ and quadrature deviation. When MMC outputs active power, $\delta_{\text{ref1}}$ is smaller than $\delta_{\text{conv1}}$, indicating $U_{\text{ref1}}\angle\delta_{\text{ref1}}$ lags behind $U_{\text{conv1}}\angle\delta_{\text{conv1}}$. Conversely, when MMC receives active power, $\delta_{\text{ref1}}$ is greater than $\delta_{\text{conv1}}$, indicating $U_{\text{ref1}}\angle\delta_{\text{ref1}}$ leads $U_{\text{conv1}}\angle\delta_{\text{conv1}}$.

Using (17a) - (23), the precise solution for RWF based on direct modulation can be derived.

*A3. Improved direct modulation*

Using $U_{\text{dcN}}$ to normalize equations (5a-5b), the definition expressions for RWF based on improved direct modulation are:

$$f_{\text{RWF},\,p} = \frac{1}{2h} - \frac{M_{\text{ref1}}}{2}\sin(\omega t + \delta_{\text{ref1}}) + \frac{M_{\text{ref2}}}{2}\sin(2\omega t + \delta_{\text{ref2}}) \tag{24a}$$

$$f_{\text{RWF},\,n} = \frac{1}{2h} + \frac{M_{\text{ref1}}}{2}\sin(\omega t + \delta_{\text{ref1}}) + \frac{M_{\text{ref2}}}{2}\sin(2\omega t + \delta_{\text{ref2}}) \tag{24b}$$

where $M_{\text{ref2}}$ is the magnitude ratios of the second-harmonic reference wave and is expressed as follows:

$$M_{\text{ref2}} = \frac{\sqrt{2}U_{\text{ref2}}}{U_{\text{dcN}}/2}. \tag{25}$$

And the factor $h$ is to evaluate deviating level of $e_{\text{d_ref}}$ from $NU_{\text{cap_ref}}$, as follows:

$$h = \frac{NU_{\text{cap_ref}}}{e_{\text{d_ref}}} = \frac{NU_{\text{capN}}}{e_{\text{d_ref}}} = \frac{U_{\text{dcN}}}{e_{\text{d_ref}}} \tag{26}$$

Obviously, in improved direct modulation, RWF will be affected by both the second-harmonic component generated by CCSC and the factor $h$ caused by the capacitor voltage control.

According to the analyses in [23] mentioned in the proposed comparison framework, the precise calculation of RWF for the improved direct modulation also needs to be divided into two categories: common mode and differential mode.

*1) Common mode of upper and lower arms*

Given that the dc component of $u_{\text{com}}$ is also equal to $U_{\text{dcN}}$, the following equation for the dc component can be established:

$$h = \frac{1 + 4c_1 I^*_{\text{ac}} M_{\text{ref1}}\sin(\delta_{\text{ref1}} + \varphi)}{1 + c_1 M_{\text{ref1}} M_{\text{ref2}} I^*_{\text{ac}}\cos(\delta_{\text{ref2}} + \varphi - \delta_{\text{ref1}})} \tag{27}$$

It is evident that the control of the capacitor voltage causes distortion in the dc component of RWF. Considering that the value of $M_{\text{ref2}}$ is significantly less than that of $M_{\text{ref1}}$, the parameter $h$ primarily depends on the numerator. When the MMC only transmits active power, $h$ approaches 1.0. However, when the MMC outputs reactive power, $h$ exceeds 1.0, and when the MMC receives reactive power, $h$ is less than 1.0.

Given the fact that the second-harmonic component of $u_{\text{com}}$ is consistently nullified due to CCSC, it is possible to derive the subsequent equations pertaining to the second-harmonic component.

$$\begin{cases} 0 = \dfrac{6c_1 M_{\text{ref1}} I^*_{\text{ac}}}{h}\cos(\delta_{\text{ref1}} - \varphi) + 2c_1 M_{\text{ref1}} M_{\text{ref2}} I^*_{\text{ac}}\cos\left(\delta_{\text{ref1}} + \varphi\right)\sin\delta_{\text{ref2}} \\ \quad -2c_1 h M^3_{\text{ref1}} I^*_{\text{ac}}\cos\left(\delta_{\text{ref1}} + \varphi\right)\cos 2\delta_{\text{ref1}} - 2c_1 M_{\text{ref1}} M_{\text{ref2}} I^*_{\text{ac}}\sin(\delta_{\text{ref2}} + \varphi + \delta_{\text{ref1}}) \\ \quad - \dfrac{2c_1 M_{\text{ref1}} M_{\text{ref2}} I^*_{\text{ac}}}{3}\sin(\delta_{\text{ref2}} - \varphi - \delta_{\text{ref1}}) + M_{\text{ref2}}\cos\delta_{\text{ref2}} \\ 0 = \dfrac{6c_1 M_{\text{ref1}} I^*_{\text{ac}}}{h}\sin(\delta_{\text{ref1}} - \varphi) - 2c_1 M_{\text{ref1}} M_{\text{ref2}} I^*_{\text{ac}}\cos\left(\delta_{\text{ref1}} + \varphi\right)\cos\delta_{\text{ref2}} \\ \quad -2c_1 h M^3_{\text{ref1}} I^*_{\text{ac}}\cos\left(\delta_{\text{ref1}} + \varphi\right)\sin 2\delta_{\text{ref1}} + 2c_1 M_{\text{ref1}} M_{\text{ref2}} I^*_{\text{ac}}\cos(\delta_{\text{ref2}} + \varphi + \delta_{\text{ref1}}) \\ \quad + \dfrac{2c_1 M_{\text{ref1}} M_{\text{ref2}} I^*_{\text{ac}}}{3}\cos(\delta_{\text{ref2}} - \varphi - \delta_{\text{ref1}}) + M_{\text{ref2}}\sin\delta_{\text{ref2}} \end{cases} \tag{28}$$

*2) Differential mode of upper and lower arms*

Considering that the fundamental-frequency component in per-unit of $u_{\text{diff}}$ is identical to actual ac output voltage $M_{\text{conv1}}\angle$

TABLE I
KEY PARAMETERS OF THE COMPARISON CASE

| Item | Value |
|---|---|
| Rated active power $P_N$ | 1250 MW |
| Rated dc-link voltage $U_{dcN}$ | 400 kV |
| Number of SMs in an arm $N$ | 200 |
| Rated SM capacitor voltage $U_{capN}$ | 2000 V |
| Equivalent interface reactance $X^*_{eq}$ | 0.25 p.u. |
| Arm reactance (converted to the valve-side) $X_{arm}{}^*$ | 0.15 p.u. |
| Leakage reactance of the interface transformer $X_T{}^*$ | 0.10 p.u |

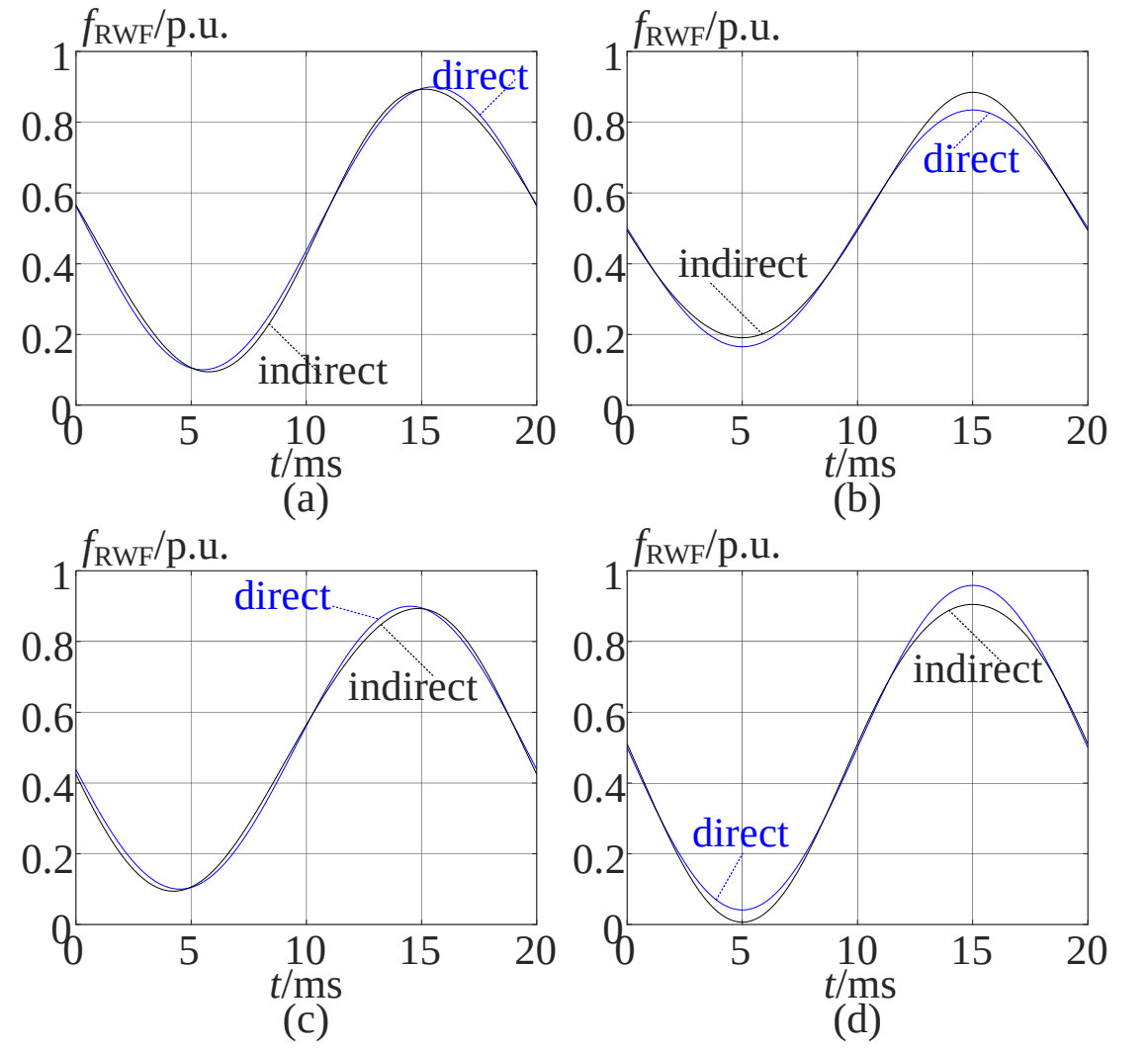

Fig. 3. Comparison of $f_{RWF}$ between direct and indirect modulation via calculation: (a) $\varphi$=-π; (b) $\varphi$=-π/2; (c) $\varphi$=0; (d) $\varphi$=π/2.

$\delta_{conv1}$, the following equations for the fundamental-frequency component can be derived:

$$\begin{cases} M_{conv1}\cos\delta_{conv1} = c_1\left[\frac{8}{h^2} - \frac{4M_{ref2}^2}{3} + M_{ref1}^2\right] I_{ac}^*\sin\varphi \\ \quad - c_1 h M_{ref1}^2 M_{ref2} I_{ac}^*\cos\left(\delta_{ref1}+\varphi\right)\cos(\delta_{ref2}-\delta_{ref1}) \\ \quad + M_{ref1}\cos\delta_{ref1} + 4c_1 M_{ref1}^2 I_{ac}^*\cos\left(\delta_{ref1}+\varphi\right)\sin\delta_{ref1} \\ M_{conv1}\sin\delta_{conv1} = c_1\left[\frac{8}{h^2} - \frac{4M_{ref2}^2}{3} + M_{ref1}^2\right] I_{ac}^*\cos\varphi \\ \quad - c_1 h M_{ref1}^2 M_{ref2} I_{ac}^*\cos\left(\delta_{ref1}+\varphi\right)\sin(\delta_{ref2}-\delta_{ref1}) \\ \quad + M_{ref1}\sin\delta_{ref1} - 4c_1 M_{ref1}^2 I_{ac}^*\cos\left(\delta_{ref1}+\varphi\right)\cos\delta_{ref1} \end{cases} \tag{29}$$

Using (24a) – (29), the precise solution for RWF of improved direct modulation can be derived.

So far, the precise quantitative analytical calculations of all the modulation techniques are comprehensive.

*B. Quantitative comparisons of RWF*

According to the proposed comparison framework, using (10a) - (16b), (17a) - (23), and (24a) - (29) in Section IV.A, the quantitative comparison of $f_{RWF}$ for direct, indirect and improved direct modulation for a specific operation point of MMC can be achieved.

In this paper, a 1250MVA/±200kV MMC is chosen as the case to compare the performance of different modulation schemes. The key parameters of the study case are presented in Table I. Meanwhile, to compare $f_{RWF}$ for different modulation techniques, the valve-side voltage $U^*_{ACV}$ is set to 0.86 p.u., $C_d$

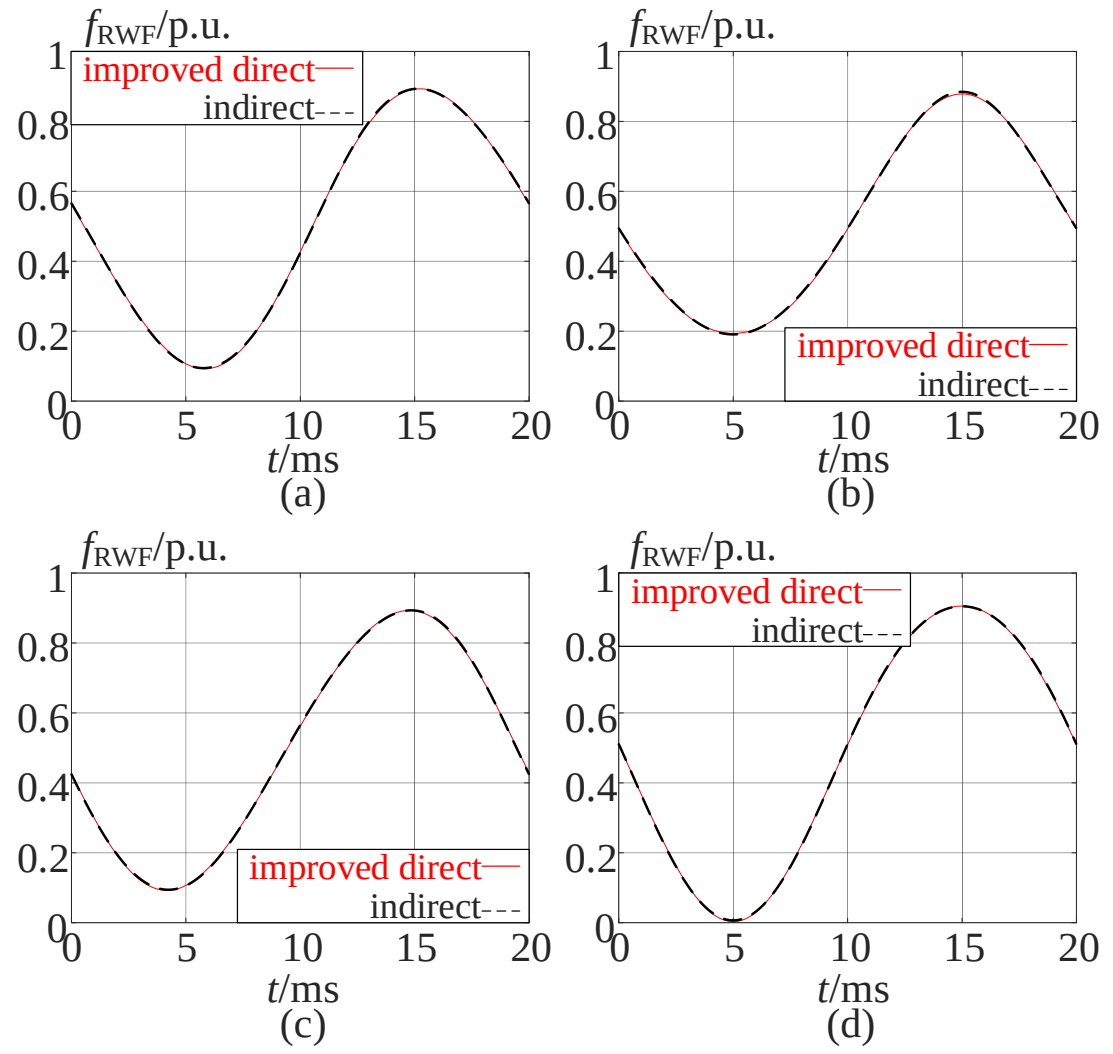

Fig. 4. Comparison of $f_{RWF}$ between improved direct and indirect modulation via calculation: (a) $\varphi$=-π; (b) $\varphi$=-π/2; (c) $\varphi$=0; (d) $\varphi$=π/2.

is fixed at 18.6 mF and output current $I^*_{ac}$=1.0 p.u.

Fig. 3 provides the comparison results of $f_{RWF}$ between direct and indirect modulation under various typical conditions. In the case of active power transmission, modulation only affects the relative phase of $f_{RWF}$. However, in the case of reactive power transmission, the situation is different. For the inductive mode in which MMC receives reactive power as depicted in Fig. 3(b), both $F_{peak}$ and $F_{valley}$ of indirect modulation are higher than those of direct modulation. Conversely, for the capacitive mode in which MMC outputs reactive power shown in Fig. 3(d), the opposite is true. Thus, the minimum linear modulation margin $\Delta F_{margin}$ in Fig.3 is located in the valley of $f_{RWF}$ in capacitive mode based on indirect modulation, indicating that once the MMC outputs more power or accesses higher valve-side voltage, it will first access overmodulation. The reason that $f_{RWF}$ of indirect modulation deviates from that of direct modulation is that there are additional dc and second-harmonic deviations in $f_{RWF}$ compared with that of direct modulation.

Fig. 4 provides the comparison results of $f_{RWF}$ between indirect and improved direct modulation under various typical conditions. The $f_{RWF}$ curves associated with distinct modulation schemes exhibit a significant degree of overlap, suggesting that two techniques are essentially interchangeable in steady-state. This new finding is clearly different from the intuitive conclusion solely based on qualitative comparison. Conventional qualitative considerations compare (10a-10b) and (24a-24b) to assert that differences exist between the two modulation techniques. However, based on the results of Fig. 4, indirect and improved direct modulation are equivalent for MMCs in steady-state. The main reason is the existence of factor $h$ caused by capacitor voltage control and the injected second-frequency reference because of CCSC.

At present, the fundamental differences among different modulation techniques in MMCs have been comprehensively examined. It is important to emphasize three points:

Firstly, direct and indirect modulation schemes have an essential difference in steady-state. On the one hand, $U_{ref1}\angle\delta_{ref1}$ in direct modulation deviates from the actual ac output voltage $U_{conv1}\angle\delta_{conv1}$ while $U_{ref1}\angle\delta_{ref1}=U_{conv1}\angle\delta_{conv1}$ in indirect modulation. On the other hand, in the normalization process, instantaneous capacitor voltages in indirect modulation are replaced by $NU_{cap_ref}$ in direct modulation.

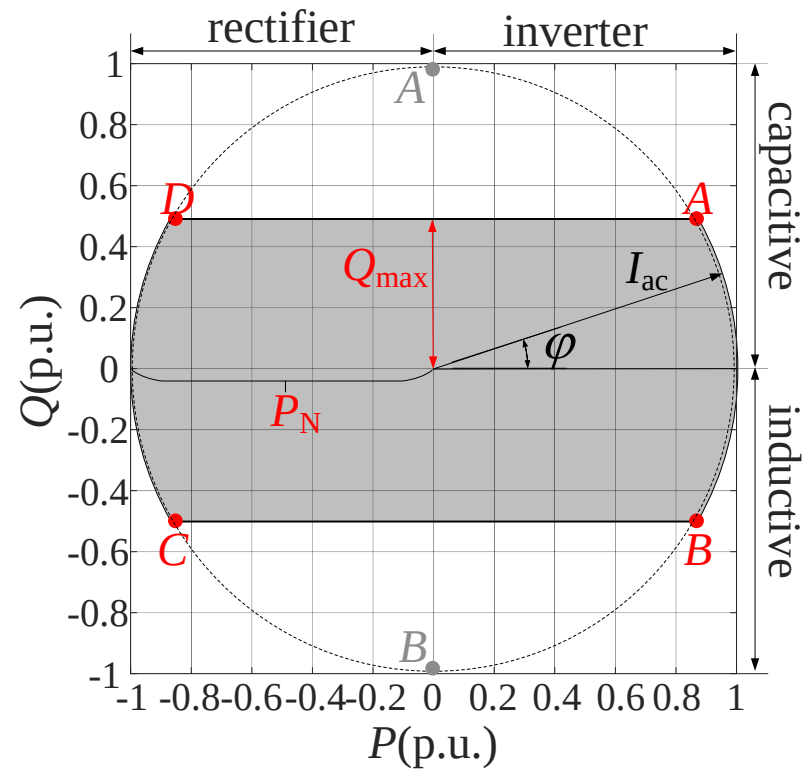

Fig. 5 Required *PQ* operating range of an MMC.

Secondly, in steady-state, indirect and improved direct modulations are equivalent for MMCs, despite their superficial differences in the implement. This means that if capacitor voltage closed-loop control and CCSC are used as additional controls for direct modulation (i.e., improved direct modulation), the corresponding steady-state performance will be similar to that based on indirect modulation.

Thirdly, direct modulation leads to dc deviation in capacitor voltage and well-known second-harmonic circulating currents while indirect and improved direct modulation introduce dc deviation and injected second-harmonic reference into RWF. And all the factors will influence modulation capability. Therefore, in next section, the comprehensive effects of different modulation schemes on the steady-state modulation capability of MMCs are compared and analyzed.

## V. Quantitative Comparisons of Modulation Capability

As shown in Fig.2, the modulation capability and operation performance of different modulation schemes for MMCs can be quantitatively compared using the metrics, including $\Delta F_{margin}$, capacitor voltages (including peak and dc values), linear *PQ* operating region based on the precise RWF obtained from Section IV. Furthermore, the effect of different modulation on parameter design can be evaluated including MS-ACV and corresponding capacitance used. The detailed analyses are as follows.

### *A. Quantitative comparisons of linear modulation margin and capacitor voltages*

Prior to quantitative comparisons of $\Delta F_{margin}$ and capacitor voltages for different modulation schemes, it is necessary to define the required operating range of the MMC. The operating range of the MMC can be characterized using the required *PQ* diagram, as depicted in Fig. 5. The diagram illustrates that active and reactive power are normalized in per-unit with respect to the rated active power. In practical applications, the required reactive power capability need not be excessively high relative to the rated active power $P_N$. Consequently, the maximum reactive power denoted as $Q_{max}$ in the *PQ* diagram

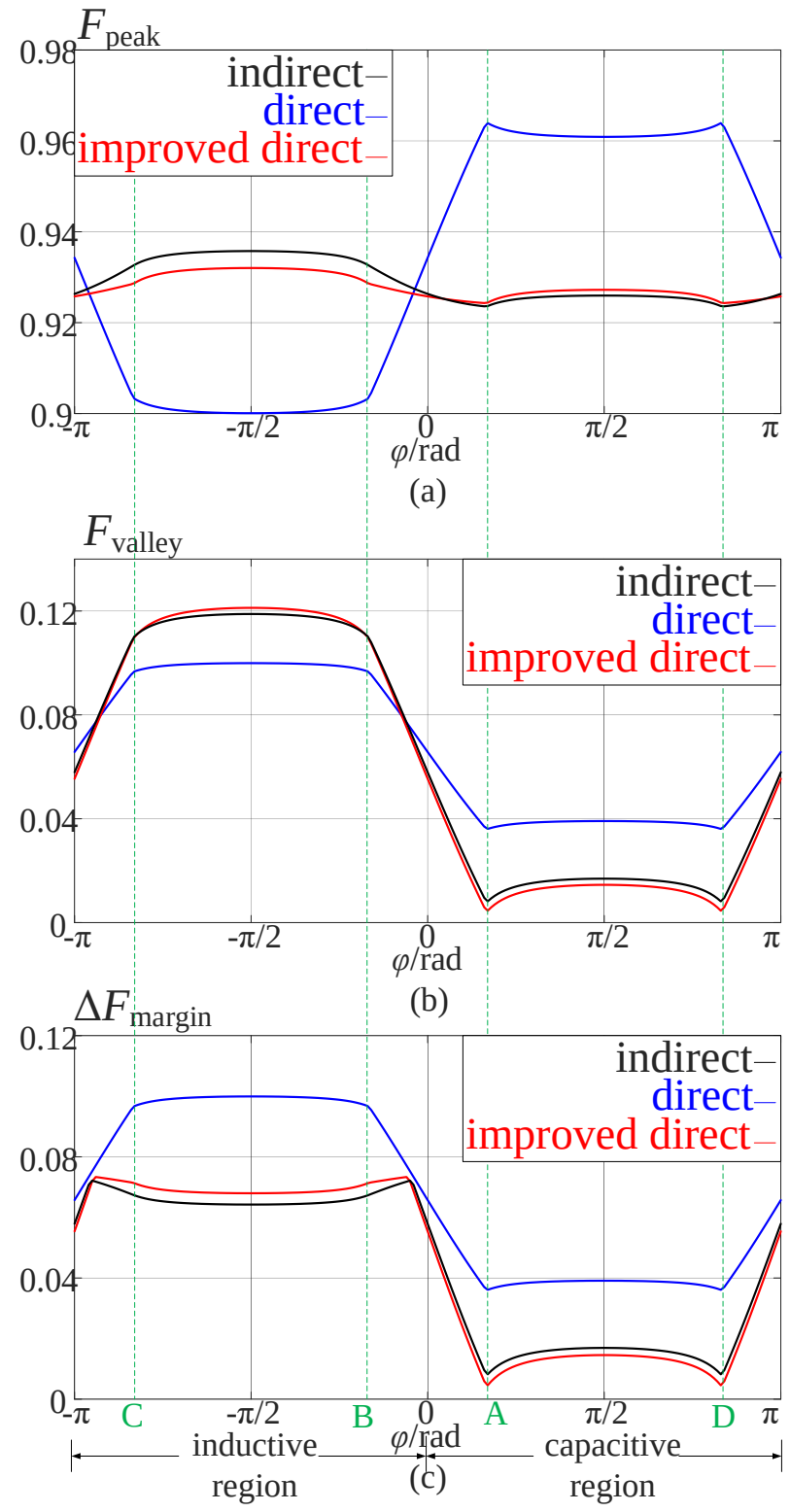

Fig. 6. Comparisons of $\Delta F_{margin}$ by scanning along the boundary of the required *PQ* operating range $U^*_{ACV}$=0.86 p.u. via calculation: (a) $F_{peak}$; (b) $F_{valley}$; and (c) $\Delta F_{margin}$

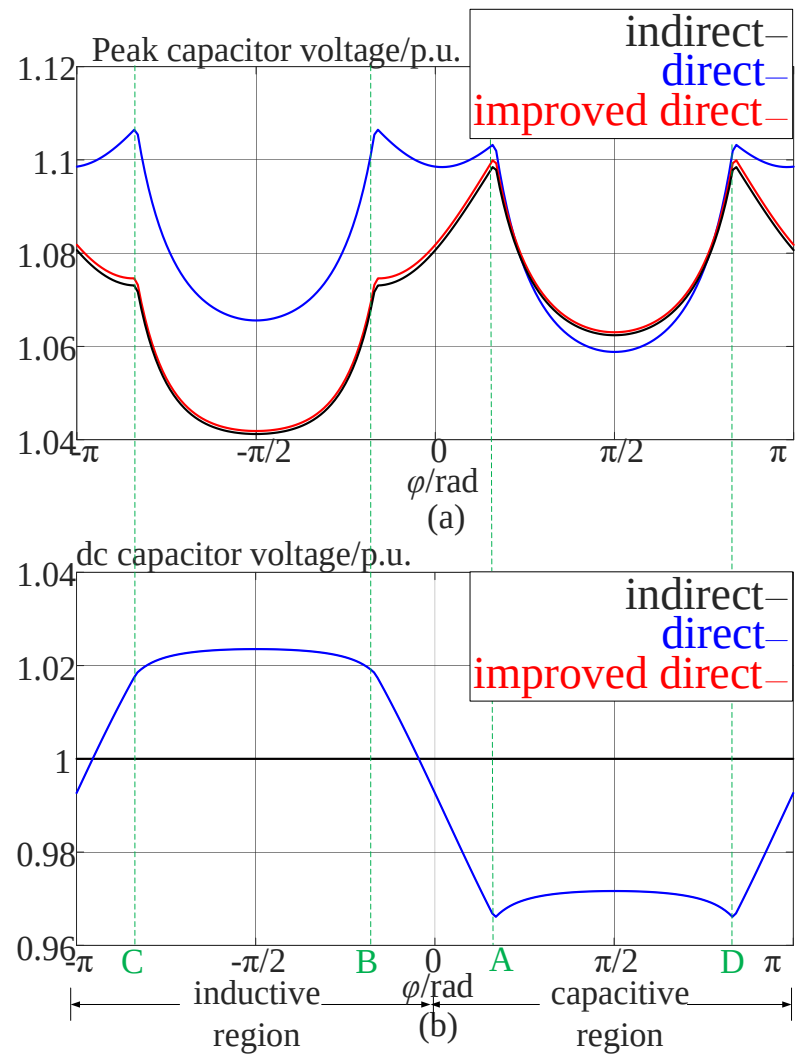

Fig. 7. Comparisons of capacitor voltages by scanning along the boundary of the required *PQ* operating range $U^*_{ACV}$=0.86 p.u via calculation: (a) peak value; (b) dc value

may vary depending on the grid requirements. Then, using the same parameters of the study case in Section III.B and $Q_{max}$=0.5 p.u., the quantitative comparison results of $\Delta F_{margin}$ and capacitor voltages for different modulation schemes can be obtained by scanning calculation for the boundary of the required *PQ* operation range, as shown in Figs. 6-7.

Fig. 6 presents detailed quantitative comparison results of $\Delta F_{margin}$ for different modulation techniques. The results reveal a significant difference between indirect and direct modulation. Specifically, when only active power is transmitted, their modulation performance is similar. However, when reactive power is transmitted, their behaviors diverge. In capacitive region where the MMC outputs reactive power (i.e., $0<\varphi<\pi$), $F_{peak}$ and $F_{valley}$ of direct modulation are both higher than those of indirect modulation. Furthermore, this gap increases as the reactive power increases. Conversely, in inductive region where the MMC receives reactive power (i.e., $-\pi<\varphi<0$), the situation is reversed. Moreover, in terms of $\Delta F_{margin}$, direct modulation consistently outperforms indirect modulation. In other words, direct modulation exhibits stronger potential than indirect modulation, indicating that it can output higher voltage. These findings clearly contradict the related argument in [21] that indirect modulation can extend the output voltage capability.

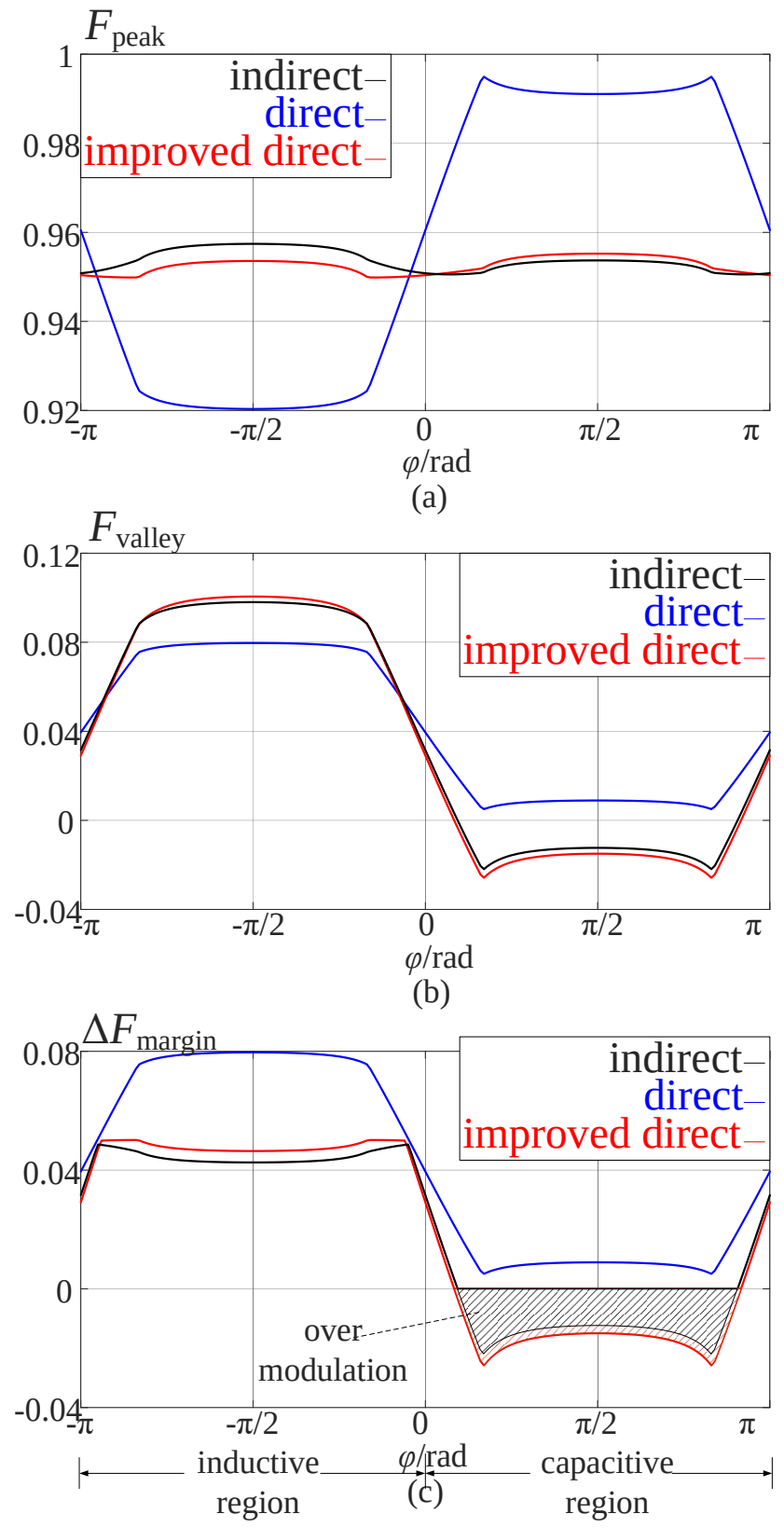

Fig. 8. Comparisons of $\Delta F_{margin}$ by scanning along the boundary of the required *PQ* operating range $U^{*}_{ACV}$=0.91 p.u. via calculation: (a) $F_{peak}$; (b) $F_{valley}$; and (c) $\Delta F_{margin}$

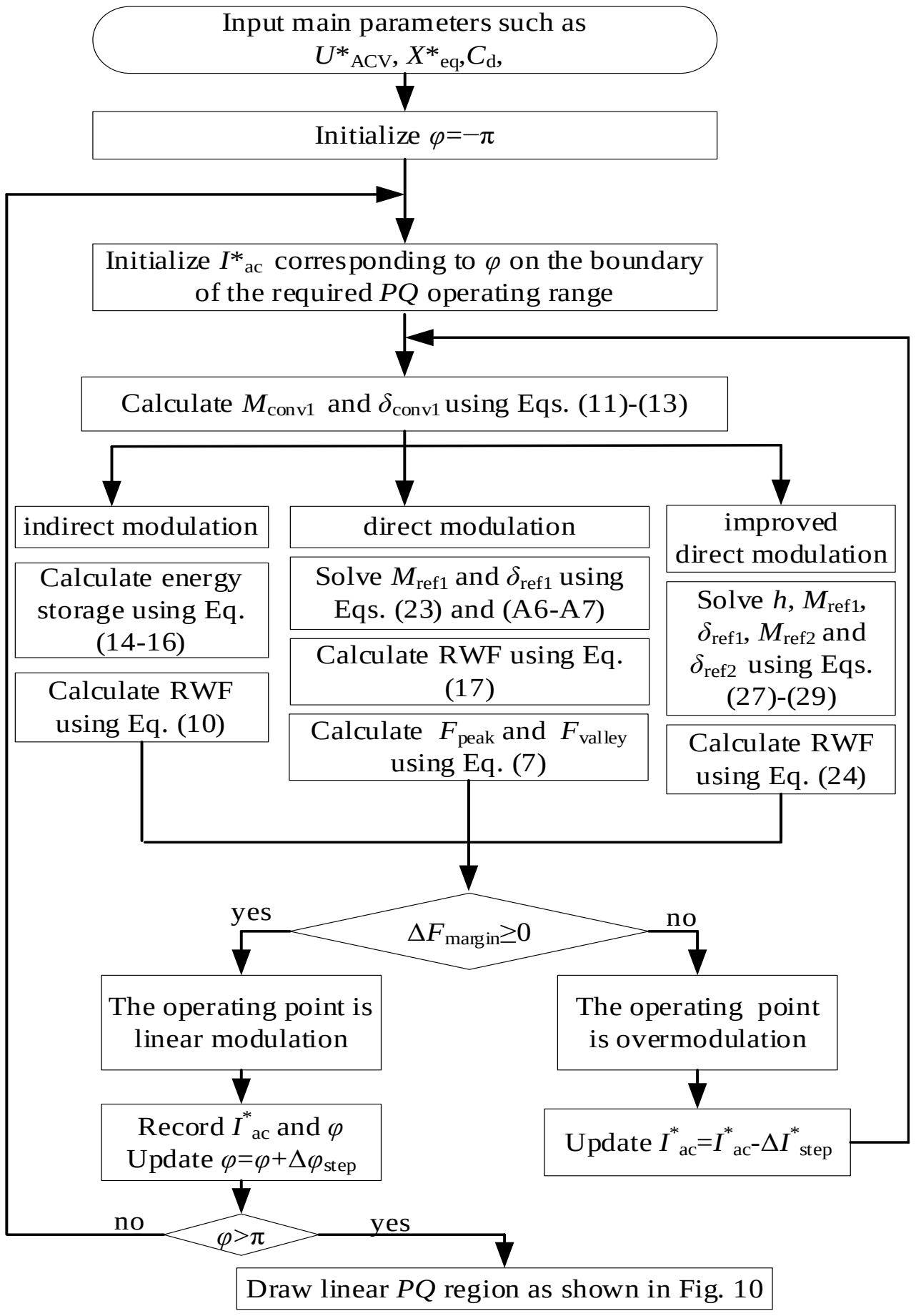

Fig. 9. Calculation flowchart for determining the linear *PQ* operating region

In addition, improved direct modulation scheme, which incorporates capacitor voltage closed-loop control and CCSC in conventional direct modulation, exhibits significantly different modulation capability compared to direct modulation. The fact that the overlapping of $F_{peak}$, $F_{valley}$ and $\Delta F_{margin}$ curves of indirect and improved direct modulation across all operating conditions indicates that the two modulation techniques are equivalent. That is, additional controls of MMCs, such as capacitor voltage closed-loop control and CCSC, decrease the linear modulation margin and output voltage potential.

Fig. 7 presents detailed quantitative comparison results of capacitor voltages (including peak and dc values) for different modulation schemes. Clearly, both the peak and dc values of capacitor voltage in improved direct modulation and indirect modulation are almost identical, while they differ from those in direct modulation. Furthermore, in most operating conditions, both indirect and improved direct modulation lead to lower peak values of capacitor voltage, and there is no dc offset in the capacitor voltage. However, a dc offset occurs in the capacitor voltage in direct modulation, which aligns with (20).

### *B. Quantitative comparisons of linear PQ operating region*

The quantitative comparisons of $\Delta F_{margin}$ in Section V.A has shown that direct modulation has stronger output potential than indirect and improved direct modulation for any linear modulation operating point. In fact, this potential can be demonstrated more intuitively when the valve-side voltage of MMC is increased. For example, when the valve-side voltage is increased to 0.91 p.u from 0.86 p.u., the value of $\Delta F_{margin}$ in different modulation schemes are depicted in Fig. 8. Obviously, as the valve-side voltage increases, $\Delta F_{margin}$ of the capacitive region based on indirect and improved direct modulation becomes less than zero, resulting in a negative region (i.e., overmodulation) as shown in Fig.8(c).

Given that Fig. 8 just considers the boundary of the required *PQ* operating range and lacks of the evaluation of linear modulation operating region, the calculation flowchart for determining the linear *PQ* operating region is proposed shown in Fig. 9. Initially, the main parameters, such as the valve-side voltage $U^{*}_{ACV}$, equivalent interface reactance $X^{*}_{eq}$, and SM capacitance $C_d$, as well as the initial power factor angle $\varphi$ and $I^{*}_{ac}$ corresponding to $\varphi$ on the boundary of the required *PQ* operating range, are input. Subsequently, the required output voltage $M_{conv1}\angle\delta_{conv1}$ is calculated, and RWF is determined according to different modulation schemes. The $F_{peak}$ and $F_{valley}$ of RWF are then calculated, and the linear modulation state is determined based on the linear modulation constraint. If the constraint is satisfied, the results including $\varphi$ and the corresponding $I^{*}_{ac}$, are recorded. And $\varphi$ is updated as $\varphi+\Delta\varphi_{step}$ to begin a new round of calculation until $\varphi=\pi$. If the constraint is not satisfied, $I^{*}_{ac}$ is updated as $I^{*}_{ac}$-$\Delta I^{*}_{step}$ to search the maximum output current corresponding to this $\varphi$. Once $\varphi$ reaches $\pi$, the calculation is concluded, and the linear $PQ$ operating region is illustrated in Fig. 10.

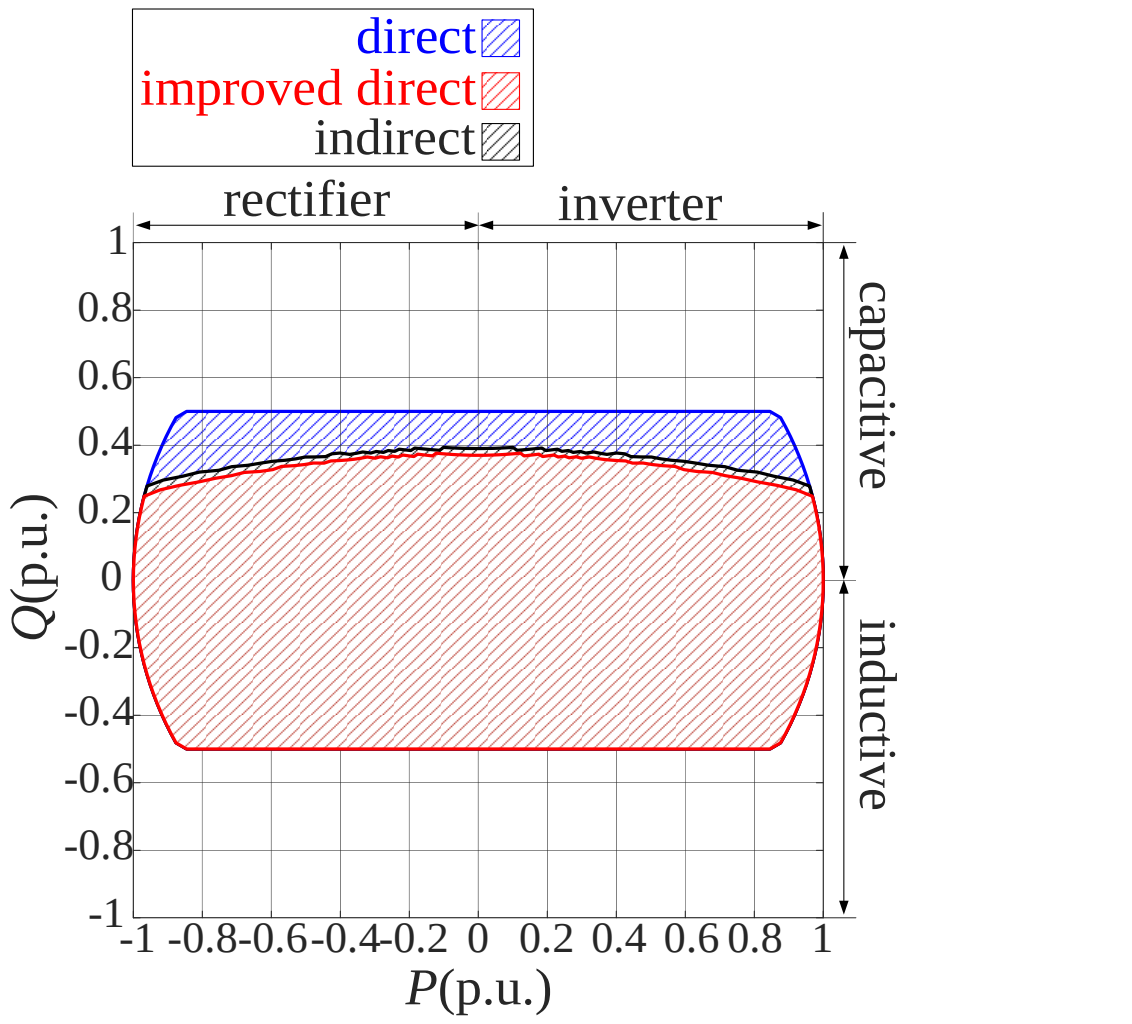

Fig. 10 Linear $PQ$ operating region comparison for different modulation via calculation

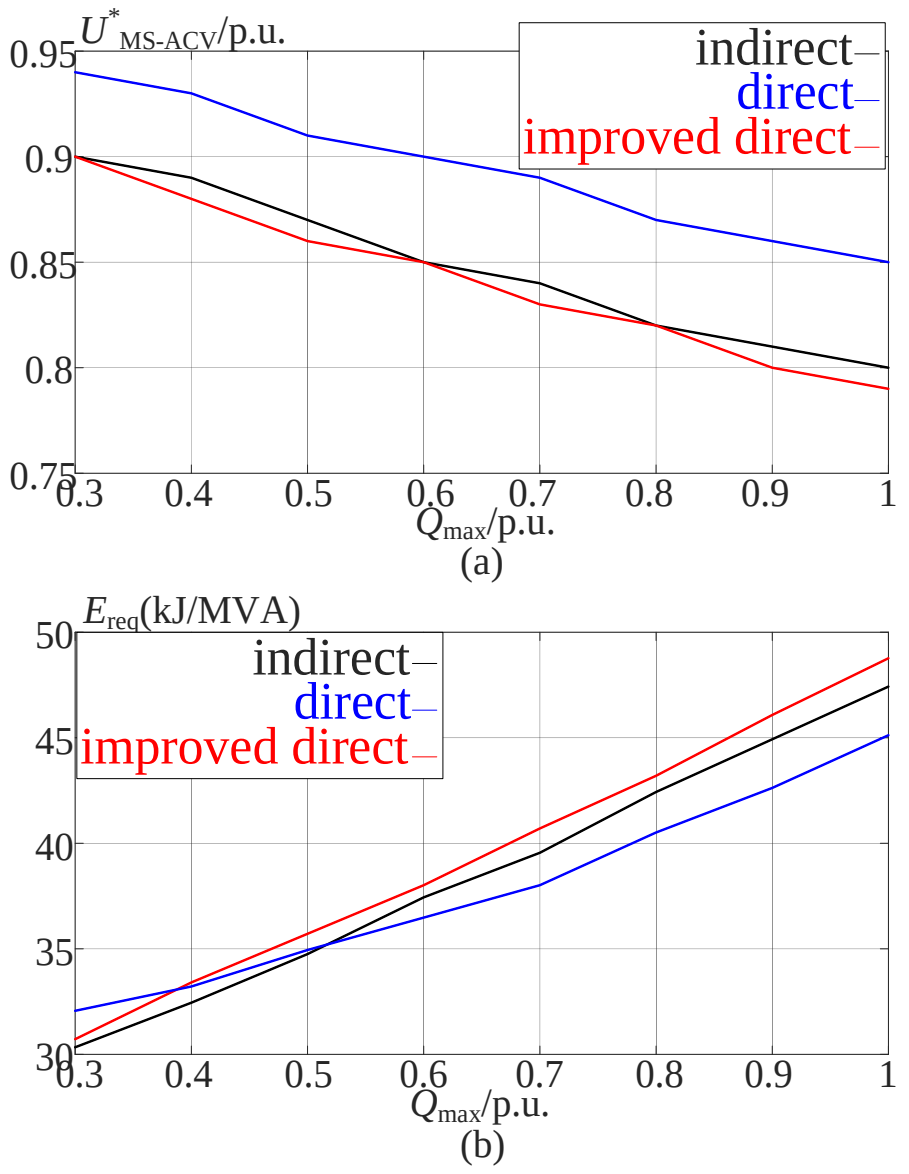

Fig. 11. MS-ACV and energy storage following the change of $Q_{max}$ via calculation: (a) $U^{*}_{MS\text{-}ACV}$, (b) energy storage $E_{req}$

Fig. 10 illustrates the linear $PQ$ operating region for different modulation schemes when the valve-side voltage $U^{*}_{ACV}$ is increased to 0.91 p.u. The linear $PQ$ operating region of direct modulation covers the entire required operating range depicted in Fig. 5, indicating that linear modulation is always satisfied by using direct modulation, which is consistent with the comparison result in Fig. 8. The linear modulation operating region of indirect and improved direct modulation almost overlap, and the region is noticeably smaller than the requirement in Fig. 5. This is because in most capacitive regions, $F_{valley}$ in indirect and improved direct modulation is lower than zero, resulting in overmodulation, which is also consistent with the comparison result in Fig. 8. And the reduction of the linear operating region is approximately 15%.

### C. Quantitative comparisons of MS-ACV and energy storage

Obviously, a higher valve-side voltage leads indirect and improved direct modulation to overmodulation and reduction of linear operating region. Thus, the maximum selectable valve-side ac voltage (MS-ACV) denoted as $U^{*}_{MS\text{-}ACV}$ which is the maximum valve-side voltage under which linear modulation is always satisfied in the whole operating range is adopted to evaluate the effects of modulation on output capability and then the selection of the valve-side voltage. Because MS-ACV is closely linked to the maximum reactive power capability $Q_{max}$ and the equivalent interface reactance $X^{*}_{eq}$, as the reactive current generates a voltage drop across the equivalent interface reactor. The value of $U^{*}_{MS\text{-}ACV}$ in different modulation techniques is investigated for variable $Q_{max}$ and $X^{*}_{eq}$=0.25 p.u., as depicted in Fig. 11(a). The value of $U^{*}_{MS\text{-}ACV}$ in direct modulation is found to be significantly greater than that of the other techniques, while indirect and improved direct modulation exhibit the similar behavior. Once this difference is ignored, the rated valve-side ac voltage will be underestimated. Additionally, Fig. 11(b) illustrates the energy storage of different modulation schemes, while the peak capacitor voltage limit is set at 1.1 p.u. The energy storage of indirect and improved direct modulation is found to be approximately equivalent and slightly lower than that of direct modulation when $Q_{max}$ is less than 0.5 p.u. However, as $Q_{max}$ increases, direct modulation emerges as the technique with the lowest capacitance used.

## VI. VERIFICATION AND EVALUATION

### A. Simulation and Experimental Verification

### A1. Simulation Verification in steady-state

To validate the proposed findings, an MMC with parameters listed in Table I is constructed in the simulation platform. The simulation results for typical operating conditions are presented in Figs. 12, which align well with Figs. 6-7. In terms of RWF, $\Delta F_{margin}$ of direct modulation consistently exceeds that of indirect modulation, indicating that direct modulation has the potential to output a larger voltage and access higher valve-side voltage, as shown in Figs. 8 and 11. On the one hand, $\Delta F_{margin}$ of capacitive region as shown in Fig. 12 (a) and (d) is lower than that of inductive region as depicted in Fig. 12 (b) and (c). On the other hand, indirect and improved direct modulation suffer a risk of overmodulation when MMC outputs maximum reactive power as depicted in Figs. 12 (a) and 12 (d), where the valley of RWF approaches the lower boundary. Besides, regarding capacitor voltage, when the MMC outputs reactive power (i.e., points A and D in Fig. 5), the peak capacitor voltage of direct modulation is approximately equal to that of indirect modulation, which is consistent with Fig. 7. However, when the MMC receives reactive power (i.e., points B and C in Fig. 5), the peak capacitor voltage of direct modulation is noticeably higher, which also aligns with the result in Fig. 7. The results show that indirect and improved direct modulation exhibit the similar behavior in RWF, capacitor voltage, which aligns with Figs. 6-8. This finding suggests that they are equivalent. In other words, direct modulation can be converted to indirect modulation through capacitor voltage control and CCSC.

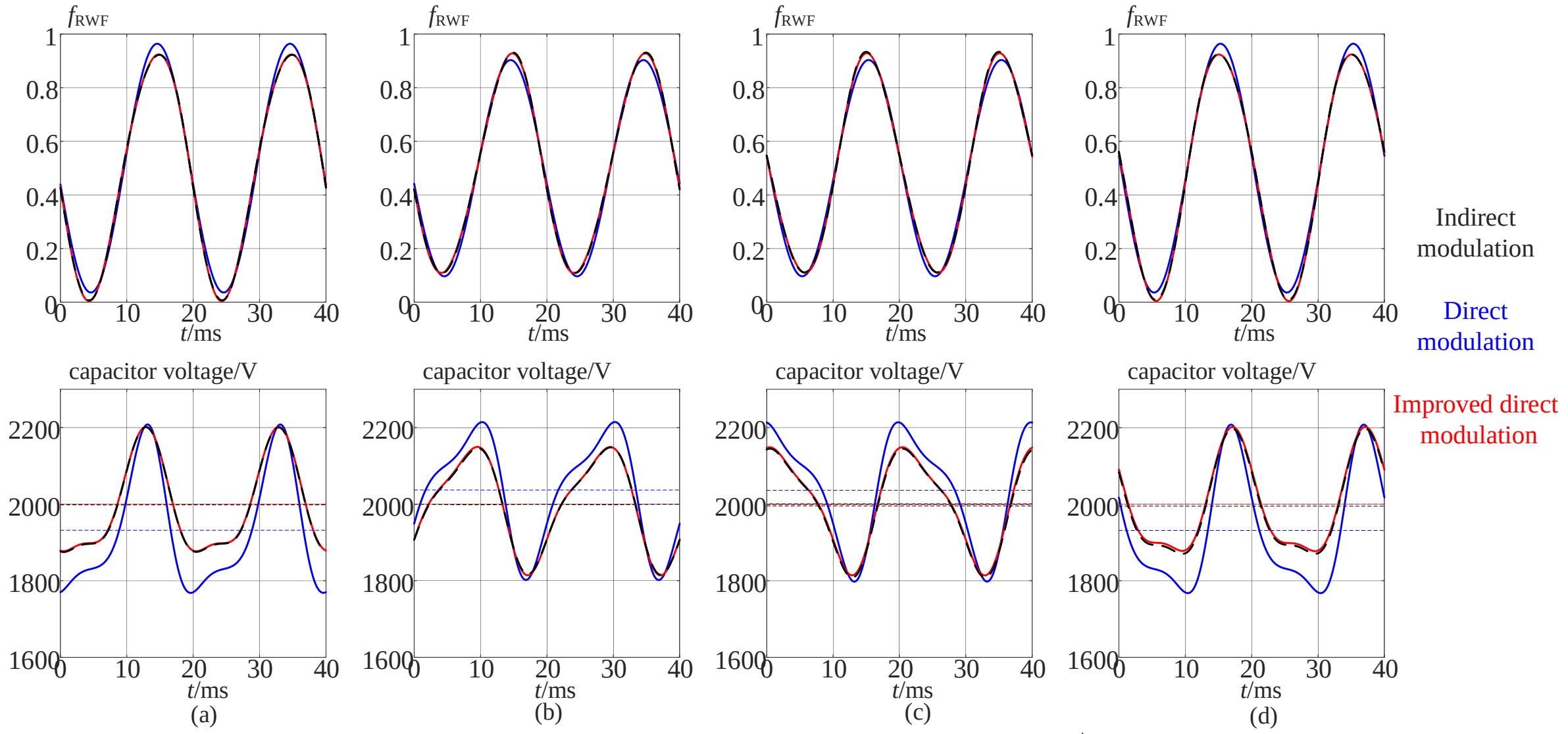

Fig. 12. RWF and capacitor voltage using different modulation schemes for typical operating conditions $U^{*}_{ACV}$=0.86 p.u., $Q_{max}$=0.5 p.u. via simulation: (a) $\varphi$=π/6 (point A in Fig. 5); (b) $\varphi$=-π/6 (point B in Fig. 5); (c) $\varphi$=-5π/6 (point C in Fig. 5); (d) $\varphi$=5π/6 (point D in Fig. 5).

*A2. Simulation Verification in transient-state*

To evaluate the transient performance of different modulation, power step conditions are simulated to verify the asymptotic stability [18-19]. And Figs. 13-14 are the transient results.

From the simulation results, several conclusions should be addressed:

1) Considering that all conditions transition to a new steady-state, it is evident that all modulations are asymptotically stable, despite their varying transient behaviors.

2) Indirect modulation effectively decouples active and reactive power of MMC, enabling precise and rapid control over power transmission. Specifically, any rapid step change in active power remains isolated and does not impact reactive power, and vice versa, ensuring optimal system performance.

3) The transient circulating current varies depending on the modulation technique, which serves to balance the arm energy storage. In the case of direct modulation, the second-harmonic circulating current shifts from its initial state to a new one. With improved direct modulation, there is a brief oscillation followed by a swift damping to zero. In contrast, indirect modulation experiences a very limited impact on the circulating current during transitions. These diverse transient behaviors present unique challenges for the design of control and protection systems.

4) In indirect modulation, the dc value of capacitor voltage is barely influenced by the step. However, improved direct modulation experiences a transient disturbance. Even more concerning is that dc deviation persists and varies with operating conditions in direct modulation due to the CVR effect [22].

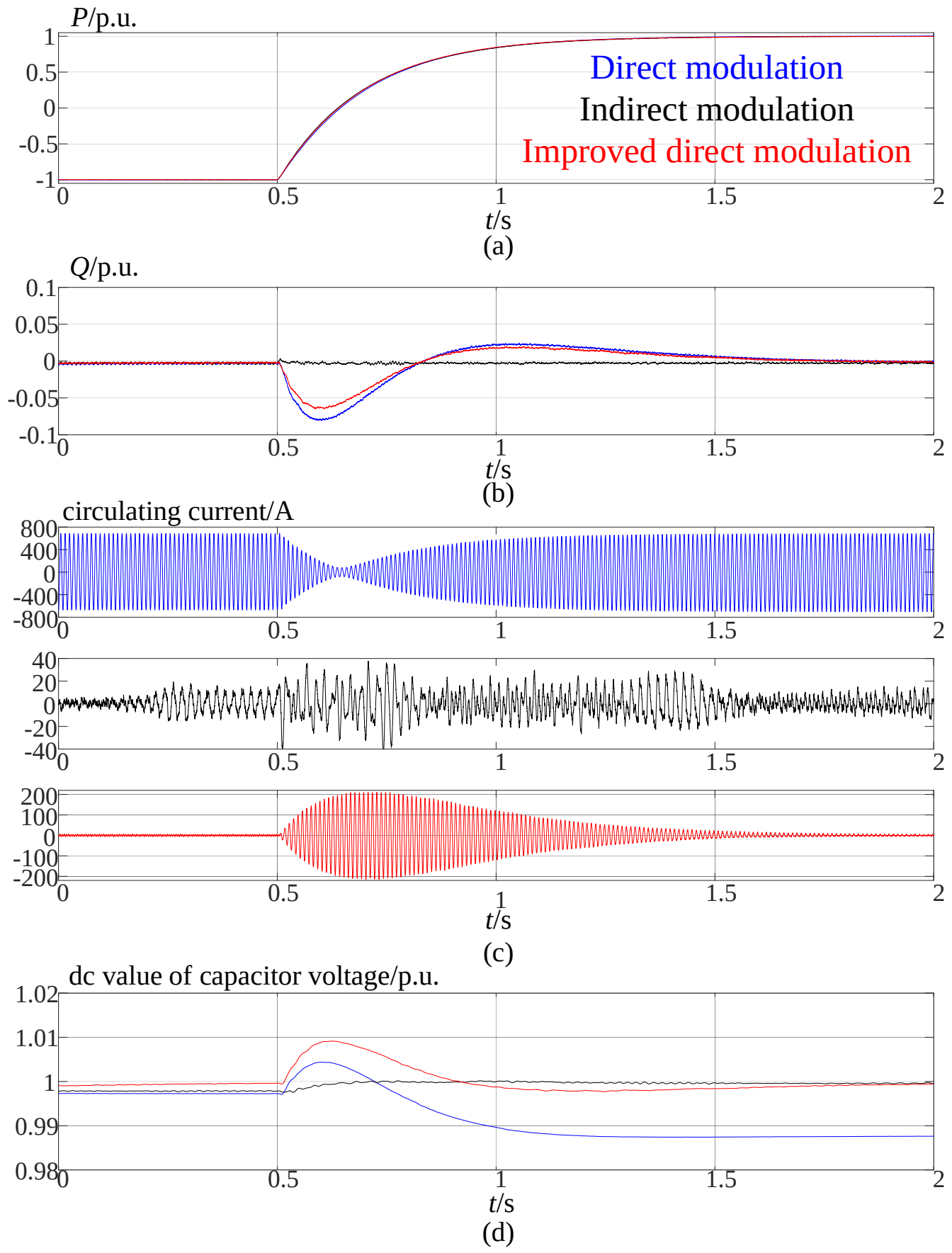

Fig. 13. Active power step from rectifier to inverter: (a) $P$; (b) $Q$, (c) circulating current, and (d) dc value of capacitor voltage

*A3. Experimental verification*

To further verify the correctness of the theoretical analysis, a laboratory prototype was constructed and tested, as shown in Fig.15. The ac side of the MMC is connected to a four-quadrant

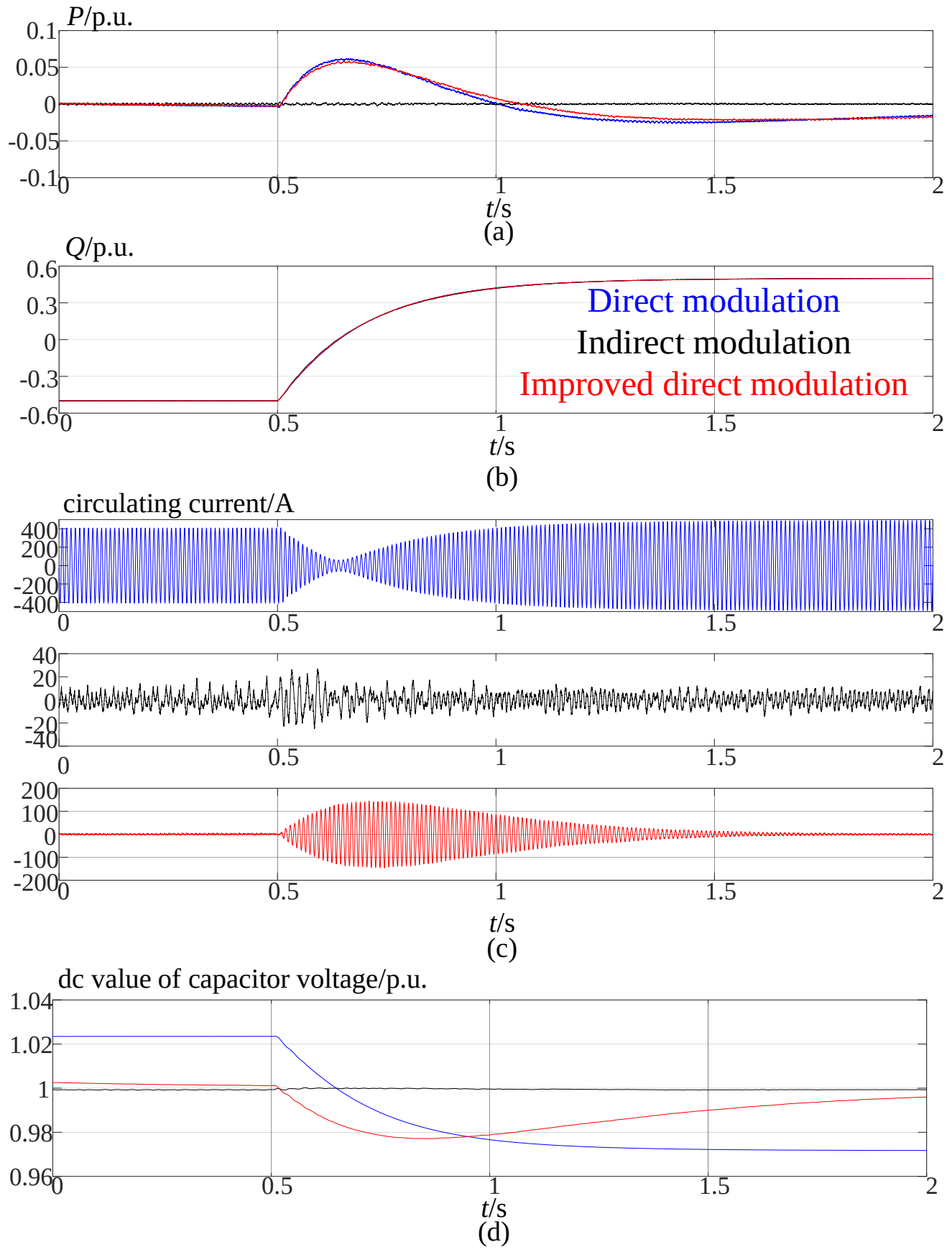

Fig. 14. Reactive power step from inductive to capacitive mode: (a) $P$; (b) $Q$, (c) circulating current, and (d) dc value of capacitor voltage

regenerative ac power source, and the dc side is connected to a programmable bidirectional dc power supply. The main parameters of the prototype are listed in Table II. The MMC prototype operated under constant active and reactive power

TABLE II

KEY PARAMETERS OF THE EXPERIMENTAL SETUP

| Item | Value |
|---|---|
| Rated active power $P_N$ | 3000 W |
| Nominal dc-link voltage $U_{dcN}$ | 300 V |
| Rated ac voltage $U_{ACV}$ | 74 V |
| Number of SMs in an arm $N$ | 4 |
| Nominal SM capacitor voltage $U_{capN}$ | 75 V |
| Arm inductor $L_{arm}$ | 13.5 mH |
| SM capacitance $C_d$ | 1.17 mF |

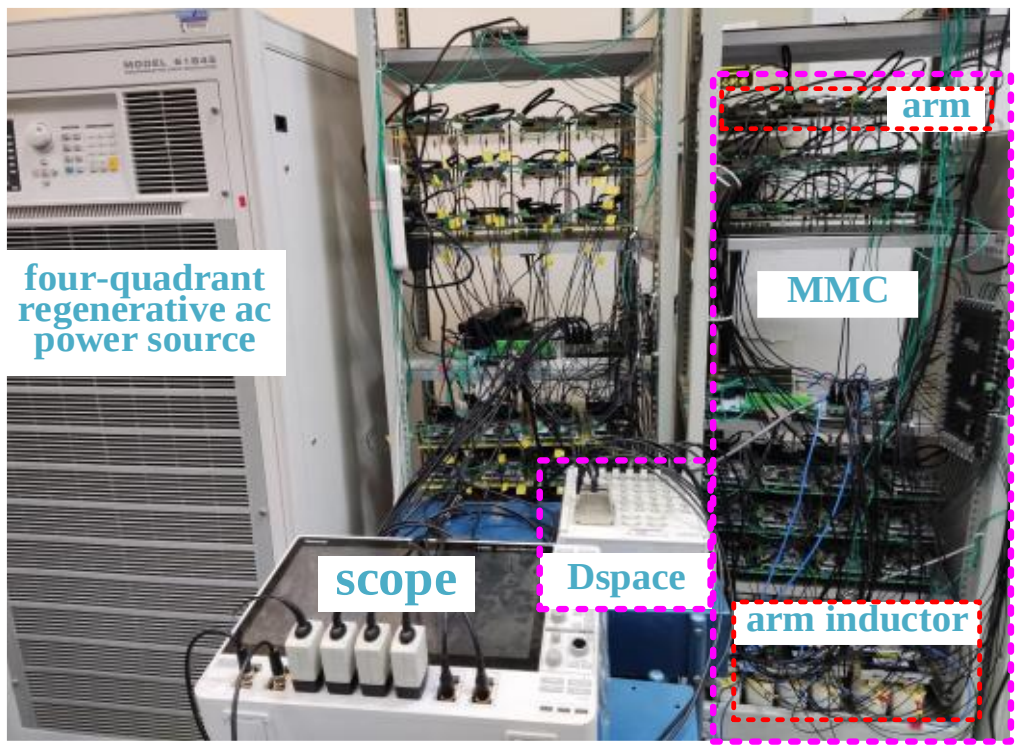

Fig. 15. Photograph of the experimental prototype

modes. Figs. 16–19 depict the experimental results for various operating conditions based on different modulation.

Using rectifier mode in Fig. 16 as an example, this study examines output voltage, capacitor voltage, arm current, and RWF including dc, fundamental- and double- frequency components. The results reveal that the three modulation schemes yield roughly equal linear modulation margin $\Delta F_{margin}$ as well as $F_{peak}$ and $F_{valley}$. And peak capacitor voltage of direct modulation is just slightly higher than those of the other modulation techniques. This suggests that in terms of linear modulation margin, the three modulation techniques perform similarly in rectifier mode, which has been indicated in Fig. 6. The above conclusions can be further corroborated in in Fig. 17, where the experimental results of the inverter mode are presented.

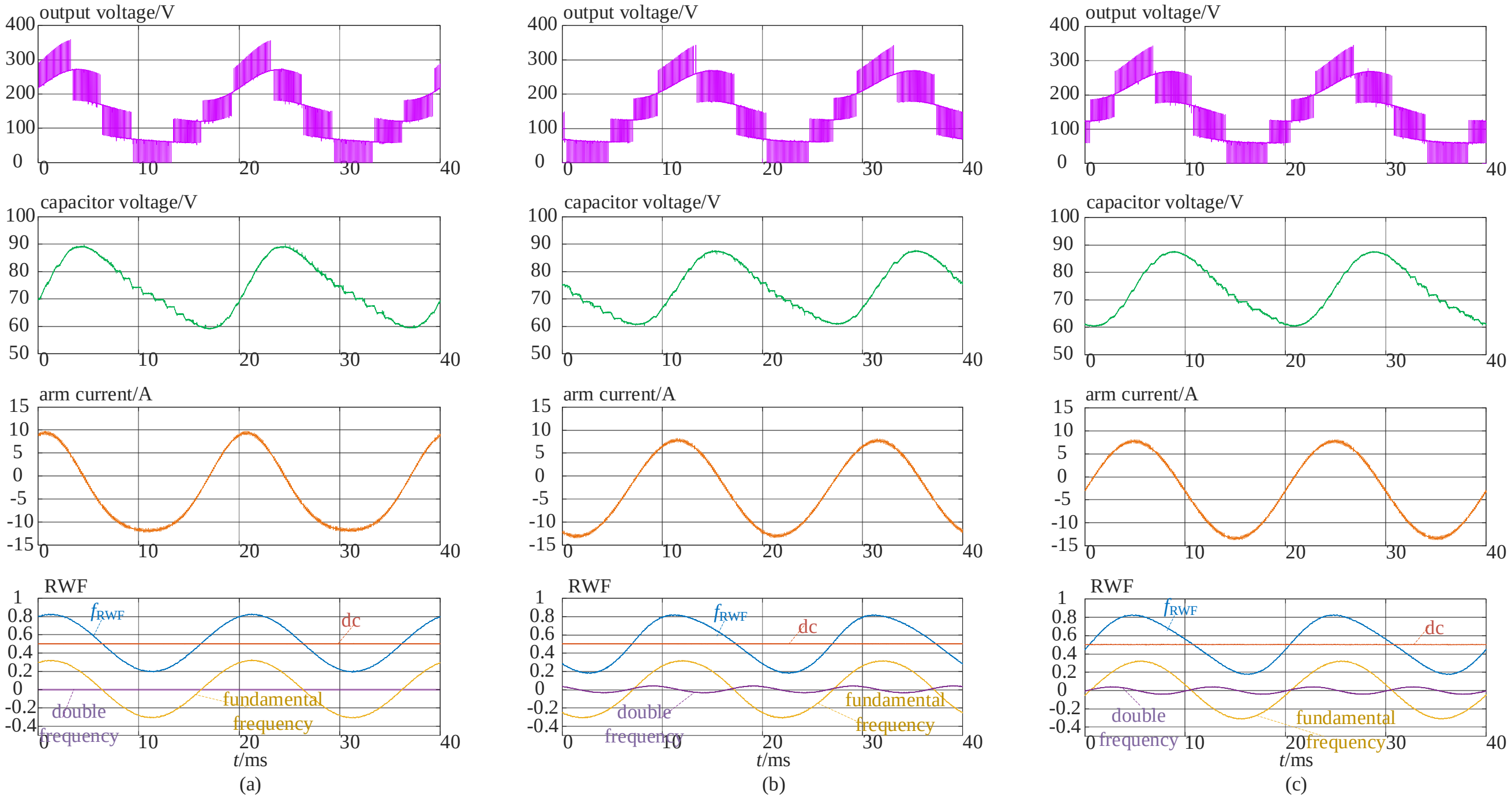

Fig. 16. Rectifier mode ($\varphi$=-π) via experiment: (a) direct modulation, (b) indirect modulation, (c) improved direct modulation.

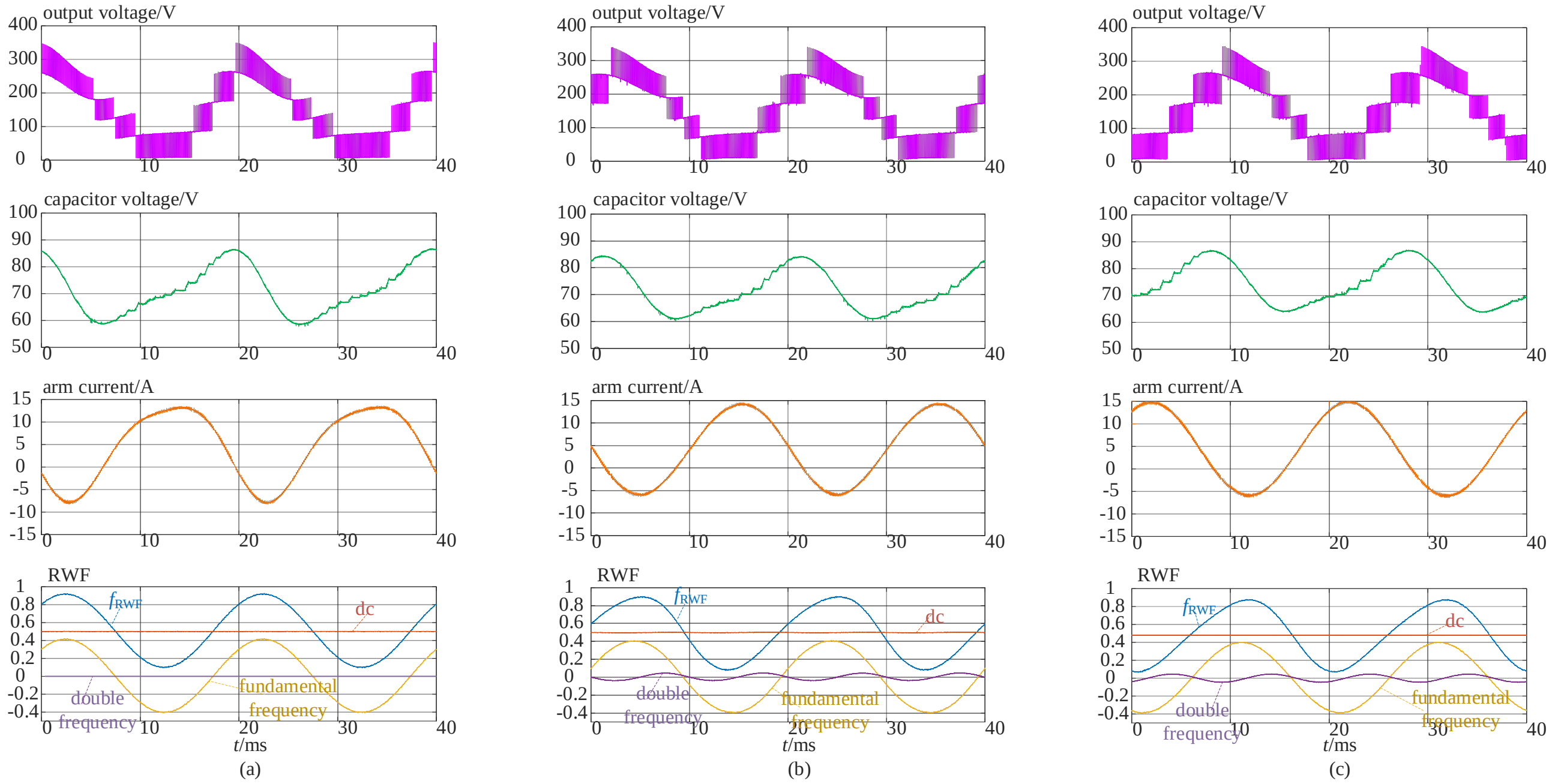

Fig. 17. Inverter mode ($\varphi$=0) via experiment: (a) direct modulation, (b) indirect modulation, (c) improved direct modulation.

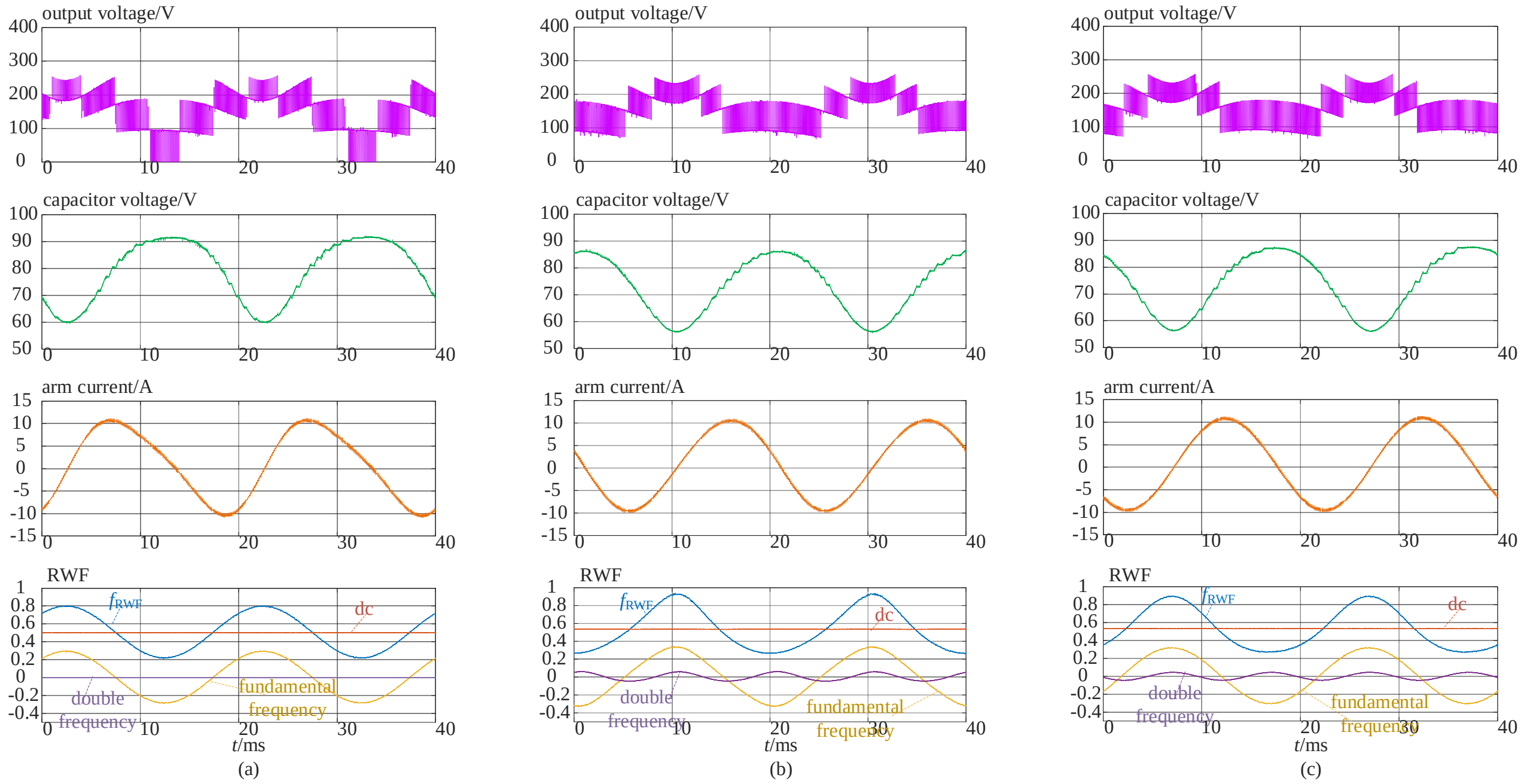

Fig. 18. Inductive mode ($\varphi$=-π/2) via experiment: (a) direct modulation, (b) indirect modulation, (c) improved direct modulation.

As for reactive power transmission, the conclusions are different. In the case of inductive mode, as depicted in Fig. 18, there is a noticeable difference in capacitor voltage, due to the positive deviation of dc value of capacitor voltage in direct modulation, resulting in a significantly higher peak capacitor voltage than those in indirect and improved direct modulation. In terms of modulation capability, the fact that dc value of capacitor voltage in indirect and improved modulation remains the same as the reference results in a higher dc reference of RWF than that encountered in direct modulation. Moreover, the peak of double-frequency reference caused by CCSC corresponds to both the peak and valley of fundamental-frequency reference. All these factors contribute to a lower linear modulation margin for indirect and improved modulation because the $F_{peak}$ is closer to the upper limit compared with direct modulation.

However, for capacitive mode, represented in Fig. 19, the situation is reversed. The observation that the dc value of capacitor aligns with the reference leads to a reduction in the dc reference of RWF for indirect and improved direct modulation. Furthermore, given that the valley of the injected double-frequency reference corresponds with both the peak and valley of fundamental-frequency reference, the linear modulation margin of indirect and improved modulation falls short of that in direct modulation. Consequently, this scenario yields the lowest $\Delta F_{margin}$, signifying a higher likelihood of overmodulation.

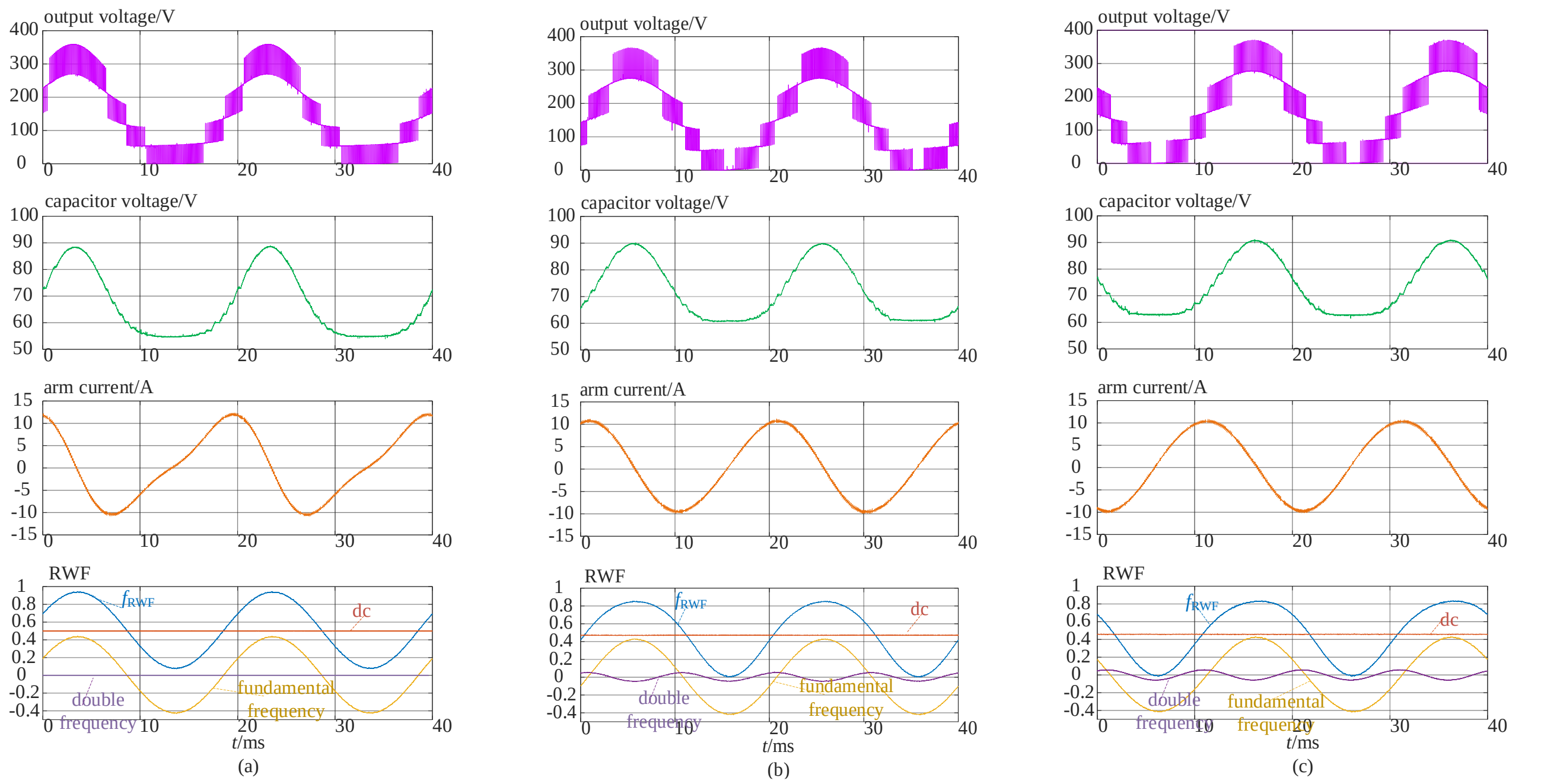

Fig. 19. Capacitive mode ($\varphi$=π/2) via experiment: (a) direct modulation, (b) indirect modulation, (c) improved direct modulation.

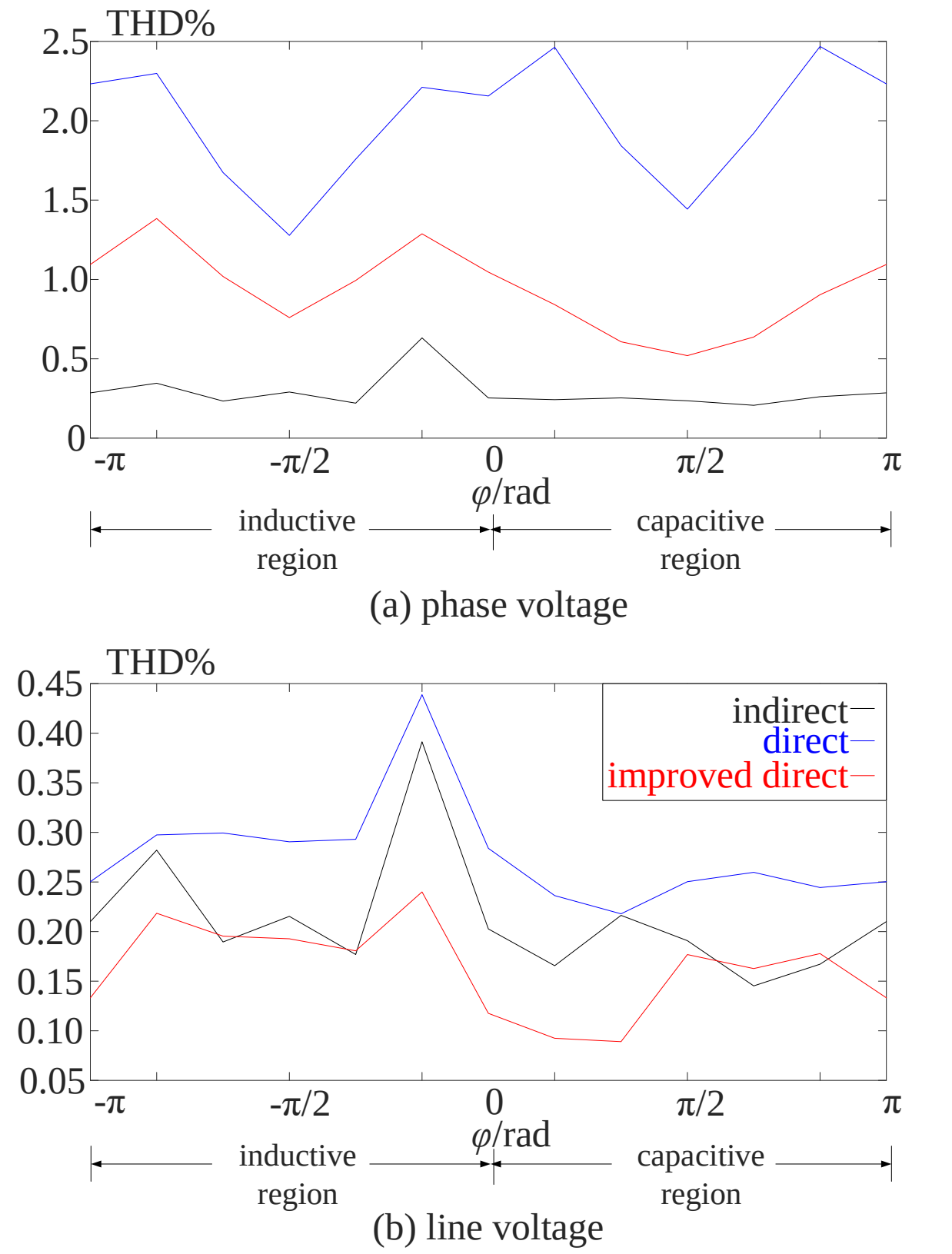

Fig. 20. Comparisons of the voltage THD by scanning along the boundary of the required *PQ* operating range via simulation

Overall, the experimental results in Figs. 16-19 show that direct modulation results in greater linear modulation margin, while indirect and improved direct modulation schemes exhibit comparable performance.

### *B. Evaluation and comparison of THD and power losses*

The ac voltage THD of the MMC has a significant impact on the power quality of the connected power grid. Meanwhile, the power losses of the MMC are closely related to the overall design. Therefore, analyzing and evaluating the voltage THD and losses is necessary for the application of the MMC. Although currently, carrier phase-shift PWM (CPS-PWM) and nearest level modulation (NLM), can both be used for MMC after the normalization process of RWF, the NLM method is more commonly used in the engineering applications of HVDC [26]-[28]. Therefore, the analyses of the voltage THD and losses in this paper is focused on the MMC utilizing NLM. The calculation method for harmonics is based on [27], while the calculation method for losses is based on [28]. According to [27] and [28], the detailed analytical calculation of the voltage THD and power losses can be obtained when the precise RWF and capacitor voltages are acquired. Then, using the same parameters of the study case in Section III.B and $Q_{max}$=0.5 p.u., the quantitative comparison results of the voltage THD and power losses for different modulation schemes also can be obtained by step-by-step scanning for the whole required *PQ* operation range, as shown in Figs. 20 and 21.

Fig. 20(a) illustrates a detailed comparison of the THD of phase voltage in the MMC for various modulation techniques. Despite the variation of THD introduced by different modulation techniques, the THD values are all very small due to the multilevel operation characteristics of the MMC. It is evident from the results that the phase voltage THD is consistently higher in the case of direct modulation compared with indirect modulation and improved indirect modulation schemes. This is because the capacitor voltage ripple results in a higher third harmonic of the phase voltage in direct modulation scheme compared to the other two schemes. Incorporating improved direct modulation scheme proves to be an effective approach in reducing the THD of the voltage. As depicted in Fig. 20(a), both the improved direct modulation and indirect modulation schemes meet the specified requirements for voltage THD outlined in IEEE Std 519-2022. According to the standard, for bus voltages greater than 161kV at the PCC, the overall THD should be less than 1.5%, and individual harmonics should be less than 1.0%. Given that the MMC is connected to the power grid with a line-to-line voltage configuration, the presence of the third harmonic components in the phase voltages will not be injected into the power grid. Therefore, the THD of the line-to-line voltage is more practically meaningful in this context. As shown in Fig. 20(b), the THD values of line-to-line voltage for all modulation techniques are significantly low. In fact, for HVDC applications, the presence of a large number of SMs provides a typical advantage in achieving low THD in MMCs, regardless of which modulation technique is employed.

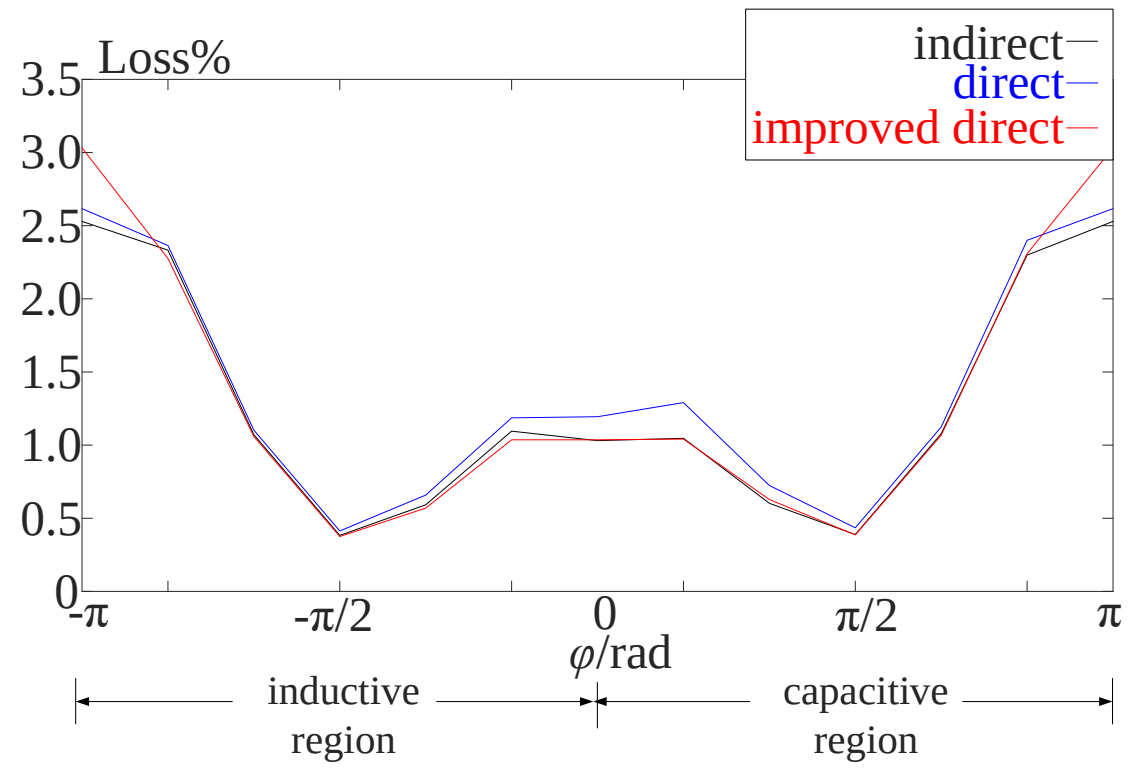

Fig. 21. Comparisons of power losses by scanning along the boundary of the required *PQ* operating range via simulation

Fig. 21 presents detailed comparison results of power losses for different modulation schemes. According to Fig. 21, the power losses associated with direct modulation are consistently higher than those of indirect modulation and improved direct modulation. However, the power losses for indirect modulation and improved direct modulation are approximately equivalent.

## VII. Conclusion

This study presents a systematic quantitative comparison of modulation techniques for MMCs based on the proposed new quantitative comparison framework, which is composed of the precise calculation of RWF and the key quantifiable metrics including linear modulation margin, capacitor voltage, and linear operating region. Furthermore, the effects of different modulation on parameter design such as the optimization of valve-side ac voltage and energy storage are assessed. Although, the framework focuses on the three modulations, it can be flexibly used for other modulations. According to the analyses, several main new conclusions can be found in this paper.

Firstly, in the terms of modulation capability, direct modulation exhibits a stronger modulation potential for MMCs. Specifically, for a certain operation point, direct modulation always has a larger modulation margin than indirect modulation. Meanwhile, for an MMC with known parameters, ac voltage output capability of direct modulation is higher and the linear *PQ* region of direct modulation is broader.

Secondly, in the steady-state, CCSC will decrease the modulation capability. This is because CCSC will inject the second-harmonic component into the RWF and then decrease the linear modulation margin. Meanwhile, it is rarely pointed out that indirect modulation eliminates the dc offset of capacitor voltage which exists in direct modulation through distorting the dc component of RWF. Similarly with CCSC, it also consumes the linear modulation margin and is negative for linear modulation. By the same token, the close-loop capacitor voltage control also decreases linear modulation capability. And the solo effect (capacitor voltage control or CCSC) on linear modulation margin is illustrated in Appendix B.

The aforementioned points require meticulous consideration for the effective operation and precise control of the MMC-HVDC project. When implementing the close-loop capacitor voltage control or CCSC in direct modulation to increase efficiency or reduce THD, it is important to consider the distortions of additional controls on RWF and then linear operating region to avoid overmodulation.

Thirdly, the effect of different modulation on parameter design including the optimization of the valve-side ac voltage and corresponding energy storage is first proposed. For any modulation, the rated valve-side ac voltage is improved as high as possible under the preset that all the points in the required *PQ* operating range satisfy linear modulation. And then the output current and capacitance used can be reduced as far as possible without overmodulation.

Lastly, the conversion condition from direct modulation to indirect modulation is revealed quantitatively. When CCSC and the close-loop capacitor voltage control are simultaneously active in direct modulation, the MMC has the same steady-state performance including RWF and capacitor voltage as that with indirect modulation, which indicates that direct modulation with additional controls, CCSC and the close-loop capacitor voltage control, is equivalent to indirect modulation in the steady-state.

Based on the above, a new design and operation for MMC-HVDC stands out: improved direct modulation can serve as a substitute for indirect modulation by achieving similar performance without the need for energy balance control and amounts of real-time communication devices in closed-loop indirect modulation or precise determination of genuine parameters and complex real-time energy storage evaluation in open-loop indirect modulation. Then, the hardware investment cost and implementation difficulty of the control system will be reduced as a result. However, though they maintain good consistency in steady-state, the differences of transient performance require special attention. In this paper, the power decoupling and transient circulating have been studied, which suggests that transient performances of different modulation may lead to different protection and safety requirement.

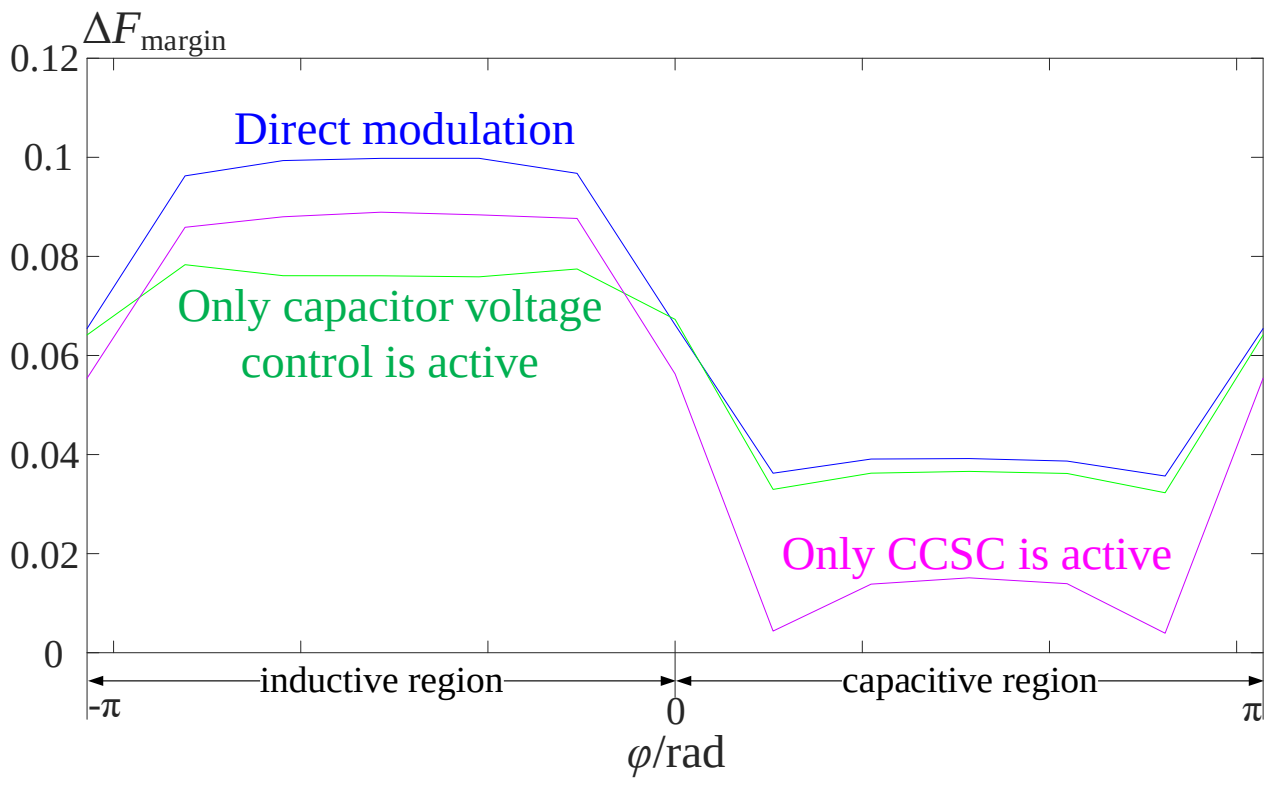

Fig. B1. Sole effect of additional control on linear modulation margin

In summary, the proposed analyses about modulation in MMC not only lies in the deepening of theory of modulation, but also provides a series of recommendations for grid operators for ensuring reasonable design and efficient operation.

## Appendix A

The fundamental-frequency, second-harmonic, and third-harmonic ripple components of capacitor voltage are calculated and expressed as follows:

$$\begin{aligned}\Delta\tilde{u}_{\mathrm{cap}(1)}(t)=&\frac{\sqrt{2}M_{\mathrm{ref1}}^{2}I_{\mathrm{ac}}}{8\omega C_{\mathrm{d}}}\cos(\varphi+\delta_{\mathrm{ref1}})\cos(\omega t+\delta_{\mathrm{ref1}})\\&-\frac{\sqrt{2}I_{\mathrm{ac}}}{4\omega C_{\mathrm{d}}}\cos(\omega t-\varphi)-\frac{\sqrt{2}M_{\mathrm{ref1}}k_{\mathrm{cir}}I_{\mathrm{ac}}}{4\omega C_{\mathrm{d}}}\sin(\omega t+\theta_{\mathrm{cir}}-\delta_{\mathrm{ref1}})\\&+\frac{\sqrt{2}M_{\mathrm{ref2}}I_{\mathrm{ac}}}{8\omega C_{\mathrm{d}}}\sin(\omega t+\delta_{\mathrm{ref2}}+\varphi)\end{aligned}\tag{A1}$$

$$\begin{aligned}\Delta\tilde{u}_{\mathrm{cap}(2)}(t)=&-\frac{\sqrt{2}M_{\mathrm{ref1}}M_{\mathrm{ref2}}I_{\mathrm{ac}}}{16\omega C_{\mathrm{d}}}\cos(\delta_{\mathrm{ref1}}+\varphi)\cos(2\omega t+\delta_{\mathrm{ref2}})\\&-\frac{\sqrt{2}k_{\mathrm{cir}}I_{\mathrm{ac}}}{4\omega C_{\mathrm{d}}}\cos(2\omega t+\theta_{\mathrm{cir}})+\frac{\sqrt{2}M_{\mathrm{ref1}}I_{\mathrm{ac}}}{16\omega C_{\mathrm{d}}}\sin(2\omega t+\delta_{\mathrm{ref1}}-\varphi)\end{aligned}\tag{A2}$$

$$\begin{aligned}\Delta\tilde{u}_{\mathrm{cap}(3)}(t)=&-\frac{\sqrt{2}M_{\mathrm{ref1}}k_{\mathrm{cir}}I_{\mathrm{ac}}}{12\omega C_{\mathrm{d}}}\sin(3\omega t+\delta_{\mathrm{ref1}}+\theta_{\mathrm{cir}})\\&-\frac{\sqrt{2}M_{\mathrm{ref2}}I_{\mathrm{ac}}}{24\omega C_{\mathrm{d}}}\sin(3\omega t+\delta_{\mathrm{ref2}}-\varphi)\end{aligned}\tag{A3}$$

where $I_{\mathrm{ac}}$ is ac output current and $\varphi$ is power factor angle. And $C_{\mathrm{d}}$ is the capacitance.

The coefficient $c_1$ is calculated as follows:

$$c_1=\frac{1}{8U_{\mathrm{ACV}}^{*}\omega E_{\mathrm{req}}}\tag{A4}$$

where $U^{*}_{\mathrm{ACV}}$ is the per-unit value of the valve side voltage $U_{\mathrm{ACV}}$ and $E_{\mathrm{req}}$ is the normalized nominal energy storage expressed as follows:

$$E_{\mathrm{req}}=\frac{\frac{1}{2}C_{\mathrm{d}}U_{\mathrm{capN}}^{2}\times N\times 6}{S_{\mathrm{N}}}\tag{A5}$$

where $S_{\mathrm{N}}$ is the nominal capacity of the converter.

The amplitude and phase of the second-harmonic circulating current can be expressed as follows:

$$k_{\mathrm{cir}}=\frac{M_{\mathrm{ref1}}\sqrt{\cos^{2}(\varphi+\delta_{\mathrm{ref1}})(3-M_{\mathrm{ref1}}^{2})^{2}+\left[3\sin(\varphi+\delta_{\mathrm{ref1}})\right]^{2}}}{\frac{X_{\mathrm{arm}}^{*}U_{\mathrm{ACV}}^{*}}{c_1}-4-\frac{8M_{\mathrm{ref1}}^{2}}{3}},\tag{A6}$$

$$\theta_{\mathrm{cir}}=\arctan\left[\cos(\varphi+\delta_{\mathrm{ref1}})(3-M_{\mathrm{ref1}}^{2}),3\sin(\varphi+\delta_{\mathrm{ref1}})\right]+2\delta_{\mathrm{ref1}}\tag{A7}$$

## Appendix B

Fig. B1 shows the sole effect of additional control (capacitor voltage control or CCSC) on linear modulation margin $\Delta F_{\mathrm{margin}}$. Obviously, either capacitor voltage control or CCSC decreases the linear modulation margin $\Delta F_{\mathrm{margin}}$. In inductive region, the reduction of closed-loop capacitor voltage is larger, while in capacitive region, the decrease of CCSC is more remarkable.